%% file: arxiv_mar2024.tex
\documentclass[fleqn,12pt]{article}
\usepackage{subcaption,graphicx}
\usepackage{xcolor}
\usepackage{amsfonts, amsmath, amsthm, amssymb,epstopdf,mathtools}
\usepackage{comment}
\usepackage{sectsty}
\usepackage[margin=2cm]{geometry}
\usepackage[round]{natbib}
\usepackage{setspace}
\usepackage{placeins, booktabs, lscape}
\usepackage{adjustbox}
\usepackage[capposition=top]{floatrow}
\usepackage{chngcntr}
\usepackage{titling}

\usepackage[toc,page,header]{appendix}

\usepackage[para,online,flushleft]{threeparttable}

\usepackage{makecell}

\usepackage{pdfpages} % to include pdf files in the document
\usepackage{bbm} % nice indicator variables

\usepackage{color}
\definecolor{Blue}{rgb}{0,0,1}
\definecolor{DarkBlue}{rgb}{0,0,0.5}
\definecolor{Red}{rgb}{1,0,0}
\definecolor{Green}{rgb}{0,1,0}
\definecolor{Yellow}{rgb}{1,1,0}
\definecolor{DarkGreen}{rgb}{0,0.4,0}
\definecolor{DarkRed}{rgb}{0.5,0,0}
\definecolor{DarkYellow}{rgb}{0.7,0.7,0}

\usepackage{hyperref}
\hypersetup{
    unicode=false,          					% non-Latin characters in Acrobat's bookmarks
    pdftoolbar=true,       						% show Acrobat's toolbar?
    pdfmenubar=true,        					% show Acrobat's menu?
    pdffitwindow=true,     						% window fit to page when opened
    pdfstartview={FitH},    					% fits the width of the page to the window
    pdfsubject={LaTeX},   						% subject of the document
    pdfproducer={Yotam Shem-Tov}, 			% producer of the document
    pdfkeywords={}, % list of keywords
    pdfdisplaydoctitle=true,					% display title instead of file name
    pdfnewwindow=true,      					% links in new window
    colorlinks=true,       						% false: boxed links; true: colored links
    linkcolor=DarkBlue,          			% color of internal links
    citecolor=DarkBlue,	        			% color of links to bibliography
    filecolor=DarkBlue,      					% color of file links
    urlcolor=blue,           					% color of external links
    citebordercolor=0 0 1,	 					% color of border around cites 0 1 0
    linkbordercolor=0 0 1,						% color of border around links 1 0 0
    urlbordercolor=0 0 1,							%	color of border around URL links 0 1 1Lafortune  (corresponding  author): 
		frenchlinks=false,    						% use small caps instead of color for links
		linktocpage=true,
    draft=false
}

\usepackage{seqsplit}
\usepackage{hyphenat}

\usepackage{tikz}
\usetikzlibrary{patterns,arrows,decorations.pathreplacing,decorations.markings, tikzmark, positioning}
\tikzstyle{startstop} = [rectangle, rounded corners, minimum width=2cm, minimum height=.8cm,text centered, draw=black]
\tikzstyle{arrow} = [thick,->,>=stealth]

\let\something\undefined %% set to false by uncommenting

\newcommand{\e}[1]{\ensuremath{\mathbb{E} \left[ #1 \right]}}

\newcommand{\indep}{\rotatebox[origin=c]{90}{$\models$}}

\newtheorem{proposition}{Proposition}

\newtheorem{ass}{Assumption}

\title{ 
\Large \textbf{
On Recoding Ordered Treatments as Binary Indicators}\thanks{
Evan K. Rose: Assistant Professor,
Department of Economics, University of Chicago, and NBER;
\href{ekrose@uchicago.edu}{ekrose@uchicago.edu}.
Yotam Shem-Tov, Assistant Professor,
Department of Economics, University of California, Los Angeles;
\href{shemtov@econ.ucla.edu}{shemtov@econ.ucla.edu}. 
We thank Denis Chetverikov, Avi Feller, Jacob Goldin, Peter Hull, John Loeser, Juliana Londo\~{n}o-V\'{e}lez, Sam Norris, Pat Kline, Rodrigo Pinto, Jonathan Roth, Andres Santos, and Lucas Zhang for helpful comments and discussions.
}
} 

\author{
Evan K. Rose   
\and 
Yotam Shem-Tov  
}
\begin{document}
\doublespacing

\begin{titlepage}
\setlength{\droptitle}{-1cm} 
\maketitle

\vspace{-6ex}
\begin{abstract}
\singlespacing
Researchers using instrumental variables to investigate ordered treatments often recode treatment into an indicator for \textit{any} exposure. We investigate this estimand under the assumption that the instruments shift compliers from no treatment to some but not from some treatment to more. We show that when there are extensive margin compliers only (EMCO) this estimand captures a weighted average of treatment effects that can be partially unbundled into each complier group's potential outcome means. We also establish an equivalence between EMCO and a two-factor selection model and apply our results to study treatment heterogeneity in the Oregon Health Insurance Experiment.
\end{abstract}

\end{titlepage}

A large literature uses instrumental variables to estimate the effects of ordered treatments such as years of education or duration of health insurance coverage \citep[e.g.,][]{angrist1991does, goldin2019health}. In these settings, it is common to recode the endogenous variable into a binary indicator for \textit{any} treatment, e.g., any college or any insurance \citep{card1995, finkelstein2012}. \cite{angrist1995} argued that doing so is a mistake. In the Local Average Treatment Effect (LATE) framework, the estimand recovered by two-stage least squares (2SLS) is the causal effect of any treatment exposure plus bias generated by intensive margin increases in treatment. This linear combination of effects is difficult to map to potential policies and may fall outside the range of treatment effects possible given the support of the outcome.

Instead, \cite{angrist1995} advocated for leaving an ordered endogenous variable unchanged. 2SLS then recovers the ``Average Causal Response" (ACR), a weighted average of effects of different doses of treatment across compliers. The ACR, however, is also difficult to interpret \citep{heckman_etal2006}. The complier groups (i.e., populations defined by their set of potential treatments) that generate it are not mutually exclusive. When treatment effects are potentially non-linear in dosage or heterogeneous---as in, for example, health responses to pharmaceuticals or the influence of unemployment duration on reemployment wages---the ACR may also differ in sign and magnitude from the effects of changes in dosage relevant for policy.

This paper investigates an assumption that significantly simplifies 2SLS analysis of ordered treatments. This restriction requires that the instruments induce units to shift from no treatment to some positive quantity but not from some treatment to more. In a study of the effects of health insurance, for example, the instruments must decrease the likelihood of being uninsured but leave the duration of coverage unchanged for individuals who would have obtained insurance regardless. In other words, the restriction requires that there are ``extensive margin compliers only" (EMCO).
In settings with one-sided non-compliance \citep[e.g.,][]{katz2001moving,kling2007experimental,heller2017thinking}---meaning individuals with one value of the instrument all receive no treatment---EMCO holds automatically. More generally, given the frequent use of recoded endogenous variables in applied research \citep[e.g.,][]{aizer2015, Bhuller_etal2018, Norris2020,finkelstein2012,Arteaga2020}, we view understanding the assumptions that can justify doing so as important.

Under EMCO, recoding an ordered treatment into an indicator is no longer a mistake. 2SLS estimates of the effect of ``any treatment" recover a weighted average of treatment effects for mutually exclusive groups of compliers. Each group is shifted to a different positive quantity of treatment from no treatment. The estimand averages the effects for each group of receiving this quantity vs. no treatment. Weights on each group are easy to recover. Moreover, under EMCO treated means for each complier group are identified, as well as the average of untreated means across all groups, allowing for a partial unbundling of the estimand. While average treatment effects for each complier group are not identified, they can be bounded. This makes it possible to test whether the data are consistent with certain hypotheses, such as that all complier groups or doses have positive treatment effects on average. Bounds can be tightened using shape restrictions motivated by the setting or economic theory, as in recent work on partial identification \citep{CSS2018}.

The power of EMCO comes from the restrictions it places on choice behavior. In the spirit of \cite{vytlacil2002}, we show that taken together the standard LATE assumptions and EMCO imply that choices are rationalized by a two-step selection model where units first decide whether to participate in treatment at all and then pick treatment levels. EMCO requires the instrument affect relative utility in the first step and not the second.
Importantly, the two-step selection process implies that at least two distinct latent factors govern treatment choices. While the presence of two factors makes marginal treatment effect analysis more complex than the single dimension usually considered \citep{heckman2010JEL}, two represent a substantial dimension reduction relative to what is generated by LATE alone, which implies selection is governed by possibly as many latent factors as levels of treatment.\footnote{For example, if treatment $D$ falls in $\{0,1,2,\dots,\bar{D}\}$, then LATE alone is equivalent to assuming there are $\bar{D}$ separate selection equations with distinct latent factors \citep{vytlacil2006}.} This dimension reduction explains why some quantities such as complier means are identified under EMCO, but not otherwise. 

In related work, \cite{eckhoff2018instrument} investigate restrictions that justify ``binarizing" an ordered treatment at a given threshold. EMCO is a special case that binarizes treatment around zero.\footnote{While we focus on the extensive margin, analogous results would apply when the instrument shifts individuals exclusively from any given level of treatment to more (e.g., from finishing high school to completing at least some college). \cite{eckhoff2018instrument}'s results, on the other hand, apply when the instrument shifts individuals exclusively from below a given level of treatment to more (e.g., from high school or less to at least some college).} Our results complement \cite{eckhoff2018instrument} by proving identification of complier means in cases where EMCO makes binarization appropriate. Moreover, we show that EMCO has clear implications for choice modeling that can help guide marginal treatment effect analysis \citep{heckman1999,heckman2005} in a setting with ordered or multiple treatments. Our results thus connect the assumptions behind binarizing approaches to the literature on the identification of complier means \citep{imbens1997,abadie2003semiparametric} and latent factor representations of LATE models \citep{vytlacil2002,vytlacil2006, heckman2018unordered}. Our results also relate to \cite{marshall2016coarsening}, who studies binarization when ruling out the exclusion restriction violations due to shifts in treatment above or below the threshold that make binarization problematic. Our assumptions, on the other hand, concern the effect of instruments on treatment but leave the effects of treatment on outcomes unrestricted. 

We conclude by examining the plausibility and implications of EMCO in \cite{finkelstein2012}'s analysis of the Oregon Health Insurance Experiment (OHIE), which randomized low-income individuals' access to Medicaid. Details of the experiment make EMCO highly likely to hold in the OHIE. To be eligible to enroll, for example, participants randomized into treatment had to be uninsured for at least six months, which implies non-compliance is one-sided. 
We use EMCO's identifying power to unpack treatment effects across complier groups. The results reveal clear patterns of intensive-margin adverse selection in the experimental data. Participants induced to remain on Medicaid the longest have the highest levels of healthcare utilization and worst self-reported health.

\section{Setting and notation} \label{sec:Setting}

Consider a setting with a single binary instrument $Z_i \in \lbrace 0, 1 \rbrace$ and a discrete, ordered treatment $D_i \in \lbrace 0,1,..., \bar{D} \rbrace$. Let $D_i(z)$ denote the treatment status of individual $i$ when $Z_i=z$. Observed treatments are $D_i = D_i(1) Z_i + D_i(0) (1-Z_i)$. Let $Y_i(d)$ denote the potential outcome of interest under treatment status $d$. Observed outcomes are $Y_i = \sum_{d=0}^{\bar{D}} 1(D_i = d) Y_i(d)$. Assume that $Z_i$ satisfies the standard assumptions of the LATE framework \citep{imbens1994} and its extension to ordered treatments \citep{angrist1995}:
\begin{ass}{\textbf{(LATE framework)}}
\label{ass:LATE_Assumptions}
\begin{align*}
& \text{(i)} \; \e{D_i|Z_i=1} > \e{D_i|Z_i=0} \quad \text{(relevance)} \\ 
& \text{(ii)} \; (Y_{i}(0), Y_i(1), \dots Y_i(\bar{D}), D_i(1), D_i(0)) \ \indep \ Z_i \quad \text{(exogeneity and exclusion)} \\
& \text{(iii)} \; D_i(1) \geq D_i(0) \quad \forall i \quad \text{(monotonicity)}  
\end{align*} 
\end{ass}

\cite{angrist1995} show that under these assumptions the Wald estimand recovers the Average Causal Response (ACR):
\begin{align}
\label{eq:acr}
\beta_{ACR} \equiv \frac{\e{Y_i|Z_i=1}-\e{Y_i|Z_i=0}}{\e{D_i|Z_i=1}-\e{D_i|Z_i=0}} = \sum_{d=1}^{\bar{D}} \omega_d \e{Y_i(d) - Y_i(d-1)|D_i(1) \geq d > D_i(0) }
\end{align}
where $\omega_d = \frac{\Pr(D_i(1) \geq d > D_i(0))}{\sum_{k=1}^{\bar{D}} \Pr(D_i(1) \geq k > D_i(0)) }$.

The ACR captures a weighted average of effects of exposure to different ``doses" of treatment (i.e., $\e{Y_i(d)-Y_i(d-1)}$) for potentially overlapping sets of compliers.\footnote{The researcher must sometimes take a stand on the correct discretization of treatment. Since time in schooling, for example, might be most accurately thought of as continuous, considering ``months" or ``years" of education is necessarily an approximation. In other settings, such as a drug trial where varying doses are administered in discrete units, no approximation is necessary.} While the ACR captures a well-defined causal parameter, it does not correspond to a clear treatment manipulation or policy counterfactual \citep{heckman_etal2006}. When treatment effects are non-linear or heterogeneous, the ACR may in fact differ in sign and magnitude from policy-relevant causal effects, such as as the average effect of a particular dosage.

Given the difficulty of interpreting the ACR, a common practice in applied research is to simply ignore the ordered nature of the treatment, recode it as binary, and interpret estimates as capturing the effects of ``any" treatment. Research on the effects of incarceration, for example, commonly uses an indicator for any prison sentence as the endogenous variable of interest \citep[e.g.,][]{aizer2015, Bhuller_etal2018, Norris2020}. \cite{angrist1995} showed that doing so may produce a biased estimator of the ACR, while results from \cite{eckhoff2018instrument} imply that the recoded endogenous variable model recovers a linear combination of effects for those shifted from zero to some treatment and those shifted from some treatment to more:

\begin{proposition}
\label{eq:any_treat_iv_linear_not_avg} Let $\beta_{recoded}$ be the Wald estimand when the endogenous variable is $1(D_i > 0)$. Then under Assumption \ref{ass:LATE_Assumptions}:
\begin{align*}
\beta_{recoded} &= 
\underset{\text{Extensive margin}}{\underbrace{\e{ Y_{i}(D_i(1)) - Y_{i}(D_i(0)) | D_i(1)>D_i(0) = 0 }}} \ 
+ \\
&\underset{\text{Intensive margin}}{\underbrace{\e{ Y_{i}(D_i(1)) - Y_{i}(D_i(0)) | D_i(1)>D_i(0) > 0 }}}  \frac{\Pr(D_i(1)>D_i(0)>0)}{\Pr(D_i(1)>D_i(0)=0)} 
\nonumber
\end{align*}
\end{proposition}
The estimator $\beta_{recoded}$ captures an unintuitive mixture of effects and thus suffers from similar interpretation issues as the ACR.\footnote{\cite{angrist1995}'s result is that $\beta_{\text{recoded}} = \beta_{\text{ACR}} \cdot (1+\kappa)$ where $\kappa \equiv \frac{ \sum_{l=2}^{\bar{D}} \Pr(D_i(1) \geq l > D_i(0)) }{ \Pr(D_i(1) \geq 1 > D_i(0)) }$. The result in Proposition \ref{eq:any_treat_iv_linear_not_avg} is a direct implication of Equations (3.7) and (3.8) in \cite{eckhoff2018instrument}, so we omit a proof. Proposition  \ref{eq:any_treat_iv_linear_not_avg} also appears in a working paper \citep{rose2018does}, which includes results from this manuscript and \cite{rose_shemtov2019does}.} Moreover, because the estimand is a linear combination and not an average, $\beta_{recoded}$ may fall outside the range of treatment effects physically possible given the support of $Y_i$. Intuitively, the primary issue is that the instrument is no longer excludable after recoding. Some individuals' outcomes may change even though their treatment status ($1(D_i > 0)$) does not. For example, when the binarized treatment is any indicator for any prison sentence, the exclusion restriction may be violated if the instrument shifts individuals from no incarceration to some prison sentence and also lengthens prison sentences for those who would have been incarcerated regardless.

These issues can be avoided if one is willing to rule out the existence of problematic complier types, an assumption we call ``extensive margin compliers only":
\begin{ass}{\textbf{Extensive Margin Compliers Only (EMCO)}} 
\label{ass:ExtensiveOnly}
\begin{align*}
 D_i(1) > D_i(0) \Rightarrow  D_i(0) = 0 \quad \forall i
\end{align*} 
\end{ass}

Assumption \ref{ass:ExtensiveOnly} requires that the instrument only causes some individuals to switch from no treatment to some, and not from some to more. This assumption will be automatically satisfied in any setting with one-sided non-compliance---i.e., where individuals with $Z_i=0$ must have $D_i=0$---but can also hold in more general settings where instruments induce some individuals to take up treatment but do not affect the relative utility of non-zero treatment levels.

EMCO implies the second term in Proposition \ref{eq:any_treat_iv_linear_not_avg} disappears and 2SLS using $1(D_i>0)$ as the endogenous variable recovers the average effect of the treatment on individuals shifted from no exposure to some positive amount.
\begin{small}
\begin{align*}
% \label{eq:any_treat_iv_causal}
\beta_{recoded} &\equiv \frac{ \e{ Y_{i} |Z_i = 1 } - \e{ Y_{i} |Z_i = 0 } }{ \e{ 1(D_i>0)|Z_i = 1 } - \e{ 1(D_i>0)|Z_i = 0 }} \\ 
&= \e{ Y_{i}(D_i(1)) - Y_{i}(D_i(0)) | D_i(1)>D_i(0) = 0 } \\
&= \sum_{d=1}^{\bar{D}} \omega_d^{r}  \e{ Y_{i}(d) - Y_{i}(0) | D_i(1)=d, D_i(0) = 0} 
\nonumber
\end{align*}
\end{small}
where $\omega_d^{r} = \frac{\Pr(D_i(1)=d, D_i(0) = 0)}{\sum_{l=1}^{\bar{D}} \Pr(D_i(1)=l, D_i(0) = 0)}$.

Our discussion so far has focused on the binary instrument case. In settings with multiple or multi-valued instruments, if EMCO applies to a comparison between two points of support in the value of the instruments, then $\beta_{recoded}$ can be estimated and interpreted treating those two points of support like the binary instrument case. If there are multiple such EMCO-compatible comparisons, the pairwise estimates can be averaged either manually or, as suggested in \cite{angrist1995}, using 2SLS. Their Theorem 2 shows that 2SLS with an ordered treatment and multiple mutually orthogonal binary instruments recovers a weighted average of ACRs for pairs of comparisons in the support of the instrument. If EMCO holds for each of these comparisons, then 2SLS using the recoded endogenous variable will thus also capture a weighted average of the $\beta_{recoded}$ estimand for each of these pairs.

Our discussion so far has also abstracted from covariates. If Assumptions \ref{ass:LATE_Assumptions} and \ref{ass:ExtensiveOnly} hold only conditional on some observed $X_i$, then $\beta_{recoded}$ can be estimated for each value in the support of $X_i$ and averaged. Alternatively, a researcher may wish to use 2SLS while controlling for a saturated set of indicators for values of $X_i$ and using the interactions of $Z_i$ and these indicators as instruments. As shown in \cite{angrist1995}, doing so produces a potentially different weighted average of the $X_i$-conditional estimates of $\beta_{recoded}$. Using linear controls that are not necessarily saturated, on the other hand, requires justifying the implicit parametric structure, as discussed in \cite{blandhol2022tsls}.

Although EMCO restricts counterfactual choices, it also implies multiple necessary conditions  must hold on the joint distribution of the outcome, treatment, and instruments. In the Online Appendix, we provide visual and formal tests of these restrictions that build on the growing literature testing instrument validity \citep[e.g.,][]{kitagawa2015test, Huber_Mellace2015, mourifie2017testing, frandsen2019judging, norris2019examiner}. These tests extend results for binary treatments from \cite{balkepearl97} and \cite{heckman2005} and are based on the observation that outcome densities must be non-negative for all complier groups. They also require that the instrument not \textit{decrease} the mass of individuals at any positive level of treatment, as is implied by EMCO's restrictions on compliance patterns. Both sets of restrictions can be tested using tools from the moment inequality literature \citep{CCK2018, bai2019practical}.

EMCO is attractive because when it holds, $\beta_{\text{recoded}}$ has a clear interpretation as a weighted average of causal effects for mutually exclusive complier populations with weights proportional to their population size. $\beta_{ACR}$ also retains a valid causal interpretation if EMCO holds, albeit one less intuitive than $\beta_{recoded}$. Under EMCO, the complier groups summed over in $\beta_{ACR}$ are simply defined by $1\{D_i(1) \geq d, D_i(0) = 0\}$ for each $d \geq 1$, and thus remain potentially overlapping.

 Treatment effect non-linearity or heterogeneity can make the magnitude and sign of treatment effects for each complier group in $\beta_{recoded}$ differ. We next show, however, that levels of each complier group's treated potential outcomes are identified and that bounds can be placed on each group's treatment effects.
 
\section{Advantages of EMCO: Complier means and bounds} \label{sec: advantages}
 
 The EMCO assumption places strong restrictions on the data generating process (DGP); however, in settings in which it is satisfied, it also provides meaningful advantages. In addition to giving $\beta_{recoded}$ a clear causal interpretation, EMCO identifies other interesting quantities.
\cite{imbens1997} and \cite{abadie2003semiparametric} show that when the treatment is binary (i.e., $D_i \in \lbrace 0,1 \rbrace$), compliers' treated and untreated mean potential outcomes are identified by 2SLS regressions using $Y_i D_i$ or $Y_i (1-D_i)$, respectively, as the outcome and $D_i$ or $(1-D_i)$ as the endogenous variable. When the treatment has multiple levels, it is tempting to try to estimate complier means using $Y_i 1(D_i=d)$ as the outcome. However, doing so without assuming EMCO yields a mixture of outcome means for multiple groups. To see why, note that the reduced form effect of $Z_i$ on this outcome is:
\begin{align} \label{eq:abadi_betaJ_discrete}
& \e{Y_i 1(D_i=d)|Z_i=1}-\e{Y_i 1(D_i=d)|Z_i=0}
\\\nonumber
& =  \underset{\text{Shifted \underline{into} $d$ from zero (extensive margin)}}{\underbrace{ \e{Y_i(d)|D_i(1)=d>D_i(0)=0} \Pr(D_i(1)=d>D_i(0)=0)}}
\\\nonumber
& + \underset{\text{Shifted \underline{into} $d$ from $D_i(0)>0$ (intensive margin)}}{\underbrace{\e{Y_i(d)|D_i(1)=d>D_i(0)>0} \Pr(D_i(1)=d>D_i(0)>0)}} \\\nonumber
& - \underset{\text{Shifted \underline{out} of $d$ to $D_i(1)>d$ (intensive margin)}}{\underbrace{\e{Y_i(d)|D_i(1)>D_i(0) = d} \Pr(D_i(1)>D_i(0)=d)}}
\nonumber
\end{align}

Hence changes in $Y_i1(D_i=d)$ due to $Z_i$ reflect individuals both moving into $D_i=d$ from multiple sources (extensive- and intensive-margin shifts) and moving into higher levels of treatment (intensive-margin shifts). Clearly Equation \ref{eq:abadi_betaJ_discrete} when rescaled by the first stage would not yield a meaningful potential outcome mean.

However, EMCO implies that $\Pr(D_i(1)=d>D_i(0)>0)$ and $\Pr(D_i(1)>D_i(0)=d)$ are both zero for $d \geq 1$. Hence treated potential outcome means for each group of ``$d$-type" compliers (i.e., individuals with $D_i(1)=d>0=D_i(0)$) are identified, as well as an average of potential outcomes under no treatment. The following proposition formalizes this claim:
\begin{proposition} \label{eq:any_treat_iv_causal_components} If Assumptions \ref{ass:LATE_Assumptions} and \ref{ass:ExtensiveOnly} hold, then:
\begin{align*}
(i) \quad \frac{\e{Y_i 1(D_i=0)|Z_i=1}-\e{Y_i 1(D_i=0)|Z_i=0}}{\e{ 1(D_i=0)|Z_i=1}-\e{ 1(D_i=0)|Z_i=0}} &= \e{Y_i(0)|D_i(1) > D_i(0)=0}
\\ &=
\sum_{d=1}^{\bar{D}} \omega_d^{r} \e{Y_i(0)|D_i(1)=d>D_i(0)=0}
\nonumber
\end{align*}
and for any $d > 0$ such that $\e{1(D_i=d)|Z_i=1} - \e{1(D_i=d)|Z_i=0} > 0$:
\begin{align*}
   (ii) \quad \frac{\e{Y_i 1(D_i=d)|Z_i=1}-\e{Y_i 1(D_i=d)|Z_i=0}}{\e{ 1(D_i=d)|Z_i=1}-\e{ 1(D_i=d)|Z_i=0}} &= \e{Y_i(d)|D_i(1)=d>D_i(0)=0} 
\end{align*}
\end{proposition}
We illustrate how these results can generate additional insights below using data from the Oregon Health Insurance Experiment. Proposition \ref{eq:any_treat_iv_causal_components} can be thought of as an extension of the results in \cite{imbens1997} and \cite{abadie2003semiparametric} for the binary case to multi-valued ordered and unordered treatments. In Appendix \ref{appen:proof_abadi_betaJ_discrete}, we present a more general version of Proposition \ref{eq:abadi_betaJ_discrete} that is analogous to Theorem 3.1 in \cite{abadie2003semiparametric} and implies that functions and means of covariates $X_i$ for each group of $d$-type compliers can also be identified, e.g.:
\begin{small}
    \begin{align}
\label{eq:compliersX}
\frac{\e{X_i 1(D_i=d)|Z_i=1}-\e{X_i 1(D_i=d)|Z_i=0}}{\e{ 1(D_i=d)|Z_i=1}-\e{ 1(D_i=d)|Z_i=0}} &= \e{X_i|D_i(1)=d>D_i(0)=0} \; \forall d > 0
% \nonumber
\end{align}
\end{small}
Unfortunately, Proposition \ref{eq:any_treat_iv_causal_components} (i) does not identify each $\e{Y_i(0)|D_i(1)=d>D_i(0)=0}$ without additional assumptions when $\omega_d^r <1 \ \forall \ d$. Intuitively, when $Z_i$ shifts individuals from $D_i(0) = 0$ to multiple positive levels of treatment, only a weighted average of untreated means across all complier groups is identified. We cannot separately identify the untreated counterfactual for each group of $d$-type compliers because there are more unknowns than equations unless there is only one complier group (which implies $\omega_d^r = 1$ for some $d$). Thus, while treated $d$-type complier means are identified, $d$-type \textit{treatment effects} are generally not: 
\begin{align*}
    & \underset{\text{Not identified}}{\underbrace{\e{ Y_{i}(d) - Y_{i}(0) | D_i(1)=d>D_i(0) = 0}}} = \\
    & \quad \underset{\text{Identified}}{\underbrace{\e{ Y_{i}(d) | D_i(1)=d>D_i(0) = 0} }} - \underset{\text{Not identified}}{\underbrace{\e{ Y_{i}(0) | D_i(1)=d>D_i(0) = 0} }} 
\end{align*}
Consequently, $\beta_{recoded}$ cannot be fully decomposed into its constituent causal components. The researcher can, however, construct bounds on $d$-type treatment effects.

Specifically, let $Y_d^0$ be the unknown quantity $\e{Y_i(0)|D_i(1)=d>D_i(0)=0}$. By the above, $Y_d^d =\e{Y_i(d)|D_i(1)=d>D_i(0)=0}$ is point identified for all $d > 0$ where $\e{1(D_i=d)|Z_i=1} - \e{1(D_i=d)|Z_i=0} > 0$. Complier group shares $\omega_d^{r}$ are likewise identified by the ratio of $\Pr(D_i = d|Z_i=1) - \Pr(D_i=d | Z_i =0)$ to $\Pr(D_i > 0|Z_i=1) - \Pr(D_i > 0 | Z_i =0)$. Bounds on $d$-type treatment effects are given by the solution to the linear program:
\begin{align}
min/max_{\{Y_d^0\}_{d=1}^{\bar{D}}}& \ Y_d^d - Y_d^{0} \\\nonumber
\textit{s.t.} \quad Y_d^0 &\in \mathrm{conv}\left(\mathcal{Y}\right) \quad \forall d > 0 \\\nonumber
\e{Y_i(0) | D_i(1) > D_i(0)} &= \sum_{d=1}^{\bar{D}} \omega_d^{r} Y_d^0 
\end{align}
where $\mathcal{Y}$ is the support of $Y_i$ and $\mathrm{conv}\left(\mathcal{Y}\right)$ denotes the convex hull of $\mathcal{Y}$.

Given that the unknown quantities $\{Y_d^0\}_{d=1}^{\bar{D}}$ are disciplined by only two sets of restrictions, these bounds are likely to be wide without further assumptions.\footnote{These bounds are not necessarily sharp. As a result of Proposition \ref{prop:generalize_Abadi2003}, for example, the full distribution of $Y_i(0)$ for the population with $D_i(1) > D_i(0)$, denoted $F_0$, is also identified. The unknown means $Y_d^0$ must also be consistent with a distribution of $Y_i(0)$ for each complier type, denoted $F_d^0$, such that $F_0 = \sum_{d=1}^{\bar{D}} \omega_d^r F_d^0$.} Imposing other shape restrictions, such as that treatment effects are decreasing in $d$, can help tighten bounds in this case. Inference can be conducted using methods that are suited to situations where the standard bootstrap fails, such as \cite{fang2019inference} or \cite{hong2020numerical}. 

Rather than bounding individual complier groups' treatment effects, it may also be interesting to test whether the data are consistent with certain joint hypotheses, such as that average treatment effects for all complier groups are weakly positive. Positive average treatment effects requires that $Y_d^0 \leq Y_d^d$ for all $d > 0$. Hence testing this hypothesis is equivalent to asking whether there exists a set of $Y_d^0$ such that:
\begin{align}
&Y_d^d - Y_d^0 \geq 0  \ \forall d > 0 \\\nonumber
&Y_d^0 \in conv(\mathcal{Y}) \quad \forall d > 0 \\\nonumber
&\e{Y_i(0) | D_i(1) > D_i(0)} = \sum_{d=1}^{\bar{D}} \omega_d^{r} Y_d^0 
\end{align}

Inference can be conducted by viewing the problem as a shape constrained generalized method of moments problem and applying methods developed by \cite{chernozhukov2015constrained}.

\FloatBarrier
\section{Implications for choice behavior} \label{sec:ChoiceBehavior}

 \cite{vytlacil2006} shows that the LATE framework laid out in Assumption \ref{ass:LATE_Assumptions}  is equivalent to a selection model defined by the following assumptions. First, treatment choices are governed by $\bar{D}$ selection equations:
\begin{flalign}
\label{eq:selectionGeneral}
1(D_i \geq d)  =  1(C^d(Z_i)-V_{i}^d \geq 0), \ \text{for} \ d \in \lbrace 1,\dots \bar{D} \rbrace
\end{flalign}
where $V_i^d$ are random variables and $C^d$ are unknown functions of the instruments satisfying $C^{d-1}(Z_i) - V_{i}^{d-1} \geq C^{d}(Z_i) - V_{i}^d \quad \forall i,d$.\footnote{As in the rest of the paper, for notational convenience we suppress implicit conditioning on observables $X_i$.}$^{,}$\footnote{Footnote 36 in \cite{rose_shemtov2019does} describes how the notation in Equation \ref{eq:selectionGeneral} relates to that in  \cite{vytlacil2006}.} Second, the instrument must induce a monotonic treatment response and be relevant, which requires that either $C^{d}(1) \geq C^{d}(0) \ \forall \ d$ and $\exists \ d \ s.t. \ \Pr\left(  C^d(1) \geq  V_i^d > C^d(0) \right) > 0$ or $C^{d}(1) \leq C^{d}(0) \ \forall \ d$ and $\exists \ d \ s.t. \ \Pr\left( C^d(1) <  V_i^d \leq C^d(0) \right) > 0$. Finally, to complete the model the instrument must be independent of both $V_i^d$ and $Y_i(d)$ for all $d$: $\left( V_i^1,\dots,V_i^{\bar{D}}, Y_i(0), Y_i(1),\dots,Y_i(\bar{D}) \right) \ \indep \ Z_i$.

While equivalent to the LATE assumptions, this selection model is difficult to work with due to the $\bar{D}$ dimensions of the unobserved heterogeneity. EMCO restricts this unobserved heterogeneity sharply. In fact, adding EMCO to the LATE framework is equivalent to assuming a two-step decision making process. In the first step, the individual chooses whether to participate or not. In the second step, the individual chooses the level of treatment. A classic example of such behavior is two-stage budgeting \citep[e.g., see][]{deaton1980economics}. This simple two-factor ``hurdle" model for treatment choices is defined by the following assumption:
\begin{ass}{\textbf{Two-factor choice model}} 
\label{eq:hurdle_simple}\\
Treatment choices are governed by
\begin{align*}
1(D_i = d) &= \begin{cases} & 1( \pi_0(Z_i) - U_i^{Ext} \geq 0 )  \ \ \text{if} \ \ d= 0 \\
& 1( \pi_0(Z_i) - U_i^{Ext} < 0 ) 1( \pi_{d+1} \leq U_i^{Int} < \pi_{d} ) \ \ \text{if} \ \ d > 0
    \end{cases} 
\end{align*} 
where $(U_i^{Ext}, U_i^{Int}) \ \indep \ Z_i$, $(U_i^{Ext}, U_i^{Int}) \sim F$ with strictly increasing marginal cumulative distribution functions, $\pi_d \geq \pi_{d+1} \ \forall d$,  $Pr(\pi_0(1) < U_i^{Ext} \leq \pi_0(0)) > 0$ or $Pr(\pi_0(0) < U_i^{Ext} \leq \pi_0(1)) > 0$, and $(Y_i(0),\dots,Y_i(\bar{D}),U_i^{Ext}, U_i^{Int}) \ \indep \ Z_i$. 
\end{ass}

Assumption \ref{eq:hurdle_simple} describes a two-equation system for treatment choices.\footnote{We use $\pi$s as notation for thresholds because $\pi_0(Z_i) = \Pr(D_i = 0 | Z_i)$ when $U_i^{Ext}$ has a marginally uniform distribution over $[0,1]$. Likewise, when $U_i^{Int}$ is marginally uniform over $[0,1]$, $\pi_{d} - \pi_{d+1} = \Pr(D_i = d | Z_i=1, D_i >0)$, as we show in Appendix \ref{appen:proofs}.} This model includes only \emph{two} latent factors: $U_i^{Ext}$, which governs the decision of whether or not to participate, and $U_i^{Int}$, which determines the level of participation. Although all threshold functions $\pi_d$ can depend on covariates $X_i$, only $\pi_0(Z_i)$ is a function of $Z_i$.  The restriction that $Pr(\pi_0(1) < U_i^{Ext} \leq \pi_0(0)) > 0$ or $Pr(\pi_0(0) < U_i^{Ext} \leq \pi_0(1)) > 0$ delivers monotonicity and relevance, while requiring that $(Y_i(0),\dots,Y_i(\bar{D}),U_i^{Ext}, U_i^{Int}) \ \indep \ Z_i$ guarantees exogeneity and exclusion.

 Proposition \ref{prop:equivalenceHurdle} formalizes the equivalence between this model and LATE plus EMCO by showing that they jointly impose the same restrictions on  behavior:

\begin{proposition}{}
\label{prop:equivalenceHurdle}
Assumptions \ref{ass:LATE_Assumptions} and \ref{ass:ExtensiveOnly} are equivalent to Assumption \ref{eq:hurdle_simple}. 
\end{proposition}

The equivalence in Proposition \ref{prop:equivalenceHurdle} is in the sense of \cite{vytlacil2002}: the selection model in Assumption \ref{eq:hurdle_simple} satisfies Assumptions \ref{ass:LATE_Assumptions} and \ref{ass:ExtensiveOnly}. But not only that, Assumptions \ref{ass:LATE_Assumptions} and \ref{ass:ExtensiveOnly} imply that one can always write down a selection model of the type in Assumption \ref{eq:hurdle_simple} that rationalizes observed and counterfactual choices. Thus, the model in Assumption \ref{eq:hurdle_simple} imposes the same restrictions on behavior as those imposed by the combination of the LATE framework assumptions and EMCO.

Because Proposition \ref{prop:equivalenceHurdle} shows that EMCO is equivalent to invoking a two-factor selection model, an implication is that ``single-index" models commonly used to reduce the dimensionality of unobserved heterogeneity in Equation \ref{eq:selectionGeneral} \citep[e.g.,][]{Dahl2002,heckman_etal2006,rose_shemtov2019does, kowalski2021reconciling} are inconsistent with EMCO except in special cases. In particular, Proposition \ref{prop:singleIndex_and_EMCO} shows that compatibility requires that if compliers are shifted to some $d > 1$, no individuals can be assigned positive treatment levels below d when $Z=0$. This requirement rules out the presence of always takers at positive levels of treatment below the maximum level of treatment obtained by treated compliers.

\begin{proposition}{}  \label{eq:SingleIndexOC_prop2}
\label{prop:singleIndex_and_EMCO} Consider the following single-index model of treatment assignment:
\begin{align*}
    1(D_i = d) = 1(\pi_{d+1}(Z_i) \leq U_i < \pi_d(Z_i)), \quad U_i \sim Uniform[0,1]
\end{align*}
where $Z_i \ \indep \ U_i$, $\pi_{d+1}(Z_i) \leq \pi_d(Z_i)$, $\pi_{d}(0) \leq \pi_d(1)$, $\pi_0(Z_i) = 1$, and $\pi_{\bar{D}+1}(Z_i) = 0$.
Then compatibility with Assumption \ref{ass:ExtensiveOnly} requires that if $\pi_d(0) < \pi_d(1)$, then $\pi_{d}(0) = \pi_{d'}(0) \ \forall \ d' \  s.t. \ 1 \leq d' < d$. In addition, compatibility can be tested by examining whether
$$\left(\e{1(D_i \geq d)|Z_i=1} - \e{1(D_i \geq d)|Z_i=0}\right)\left(1-\e{1(D_i \geq d)|Z_i=0} - \e{1(D_i = 0)|Z_i=0}\right) = 0$$ holds for all $d > 1$.
\end{proposition}

The testable implication contained in Proposition \ref{eq:SingleIndexOC_prop2} is straightforward. Single-index compatibility with EMCO requires that for levels of treatment greater than one either $\Pr(D_i(1) \geq d) = \Pr(D_i(0) \geq d)$, so that no compliers are shifted to levels of treatment $\geq d$ (implying the first parenthetical term is zero) or $\Pr(1 \leq D_i(0) < d) = 0$ (implying the second parenthetical term is zero).

Instruments that satisfy EMCO-like behavioral restrictions are commonly used in economics. For example, the need for such instruments arises naturally in labor economics when researchers studying wages seek to correct for the choice to work at all in the style of \cite{gronau1974wage} and \cite{heckman1974shadow}. \cite{mulligan2008selection} use these techniques to estimate the influence of the changing composition of women in the labor force on the gender wage gap. Their instrument---a mother's number of children aged zero to six interacted with marital status---must impact the decision to work but not labor supply among mothers already working. Another example comes from \cite{card2005estimating}, whose strategy to estimate the wages of individuals induced into working by a time-limited earnings subsidy can be justified by an EMCO restriction on the effects of the subsidy on labor supply.

\subsection{Implications of EMCO for marginal treatment effect analysis}

The inconsistency of EMCO with single-index models makes marginal treatment effect (MTE) analysis \citep{heckman1999,heckman2005} more complex. While two factors represent a substantial dimension reduction relative to the $\bar{D}$ implied by LATE alone, modeling treatment effect heterogeneity in two dimensions is significantly more challenging than the single dimension considered in the MTE literature \citep[e.g.,][]{heckman2010JEL} and in the ordered treatment cases studied in \cite{rose_shemtov2019does}.

One simplifying assumption that would restore the validity of standard MTE tools is that potential outcomes are not affected by latent factors that govern selection along the intensive margin:   
\begin{align}
    \e{Y_i(d)|U_i^{Ext}, U_i^{Int}} = \e{Y_i(d)| U_i^{Ext}}
\end{align}  
However, this assumption precludes selection into treatment based on gains and levels along the intensive margin except through correlation between $U_i^{Ext}$ and $U_i^{Int}$.

An alternative approach is to model potential outcomes as functions of both $U_i^{Ext}$ and $U_i^{Int}$. The researcher can then estimate or bound other treatment effects of interest (such as an average treatment effect) consistent with the moments identified under EMCO. Specifically, let $m_d(u_1,u_2) = \e{Y_i(d) | U_i^{Ext} = u_1, U_i^{Int} = u_2}$ represent treatment response functions. $d$-type compliers' potential outcome means are given by:
\begin{small}
\begin{align}
    & \e{Y_i(d)|D_i(1)=d>0=D_i(0)} = 
      \\ &\quad = \int_{\pi_{d+1}}^{\pi_d} \int_{\pi_0(0)}^{\pi_0(1)}m_d(u_1,u_2)dF(u_1,u_2) / \int_{\pi_{d+1}}^{\pi_d}\int_{\pi_0(0)}^{\pi_0(1)}dF(u_1,u_2) \nonumber
\end{align}
\end{small}
The researcher can then pick explicit functional forms for treatment response functions or flexibly approximate them in the style of \cite{mogstad2018using} and \cite{marx2020sharp}. Each identified complier mean serves to discipline these functions. 

\FloatBarrier
\section{Empirical application: The effects of health insurance} \label{sec:Applications}

In 2008, a group of low-income adults in Oregon were randomly given the opportunity to apply for Medicaid. \cite{finkelstein2012} use this experiment---dubbed the Oregon Health Insurance Experiment (OHIE)---to study the effects of access to Medicaid on health care utilization and financial and physical well-being. They find that insurance increases both primary and emergency care utilization, lowers some health care expenditures, and increases self-reported physical and mental health. 

To analyze the experiment, the researchers primarily use 2SLS specifications with ``ever on Medicaid" as the endogenous variable. However, they note that the treatment in their setting---duration of Medicaid coverage---is continuous, and argue that coding the treatment as the ```number of months on Medicaid' may be more appropriate than `ever on Medicaid' where the effect of insurance on the outcome is linear in the number of months insured." While a continuous endogenous variable would be appropriate regardless of linearity in the effects of insurance, a binary endogenous variable for ``any Medicaid" is also appropriate in this setting because EMCO is highly likely to hold for institutional reasons. The OHIE randomized admission into the Oregon Health Standard Plan, which had been closed to enrollment since 2004. Subjects lotteried into treatment were able to enroll and remain on the plan so long as they were eligible, which required being uninsured for at least six months prior to enrolling and ineligible for other public health insurance programs.\footnote{Candidates also had to be 19–64 years of age, US citizens or legal immigrants, have income under 100 percent of the federal poverty level, and possess assets of less than \$2,000.} Individuals who would have obtained some Medicaid in the control group therefore could not increase their duration of coverage if lotteried into treatment, since they would be ineligible.\footnote{It is possible some subjects would have obtained Medicaid through other means if not lotteried into treatment. However, the data in this case are consistent with EMCO holding nevertheless. Moment inequality tests of EMCO's restrictions (see Appendix \ref{sec: testable} for details) support its applicability to the OHIE.}

\ifdefined\something
%   True
    \begin{figure}[!ht]{}
    \caption{Visual Tests of EMCO in \cite{finkelstein2012}'s Analysis of the OHIE}  
    \label{fig:ohie1}
    \begin{tabular}{cc}
    \includegraphics[scale=0.56]{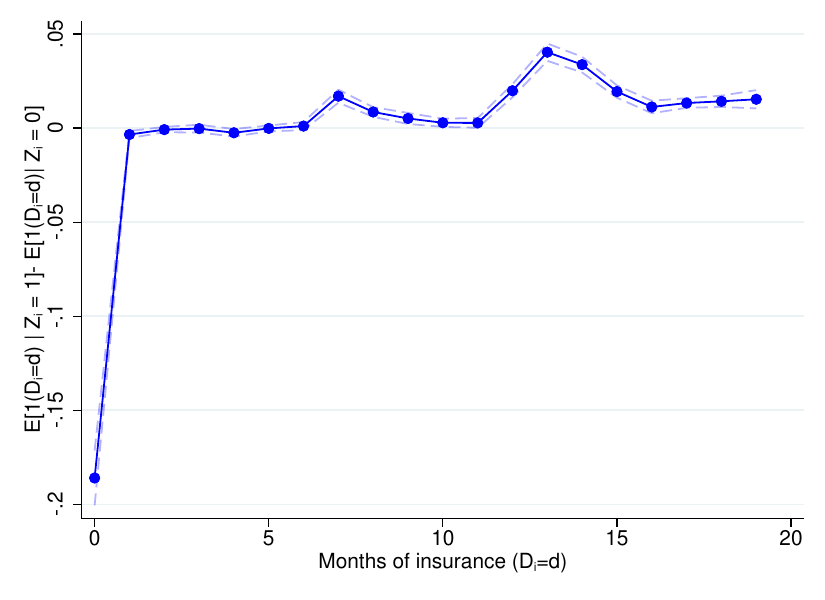}
    &
    \includegraphics[scale=0.56]{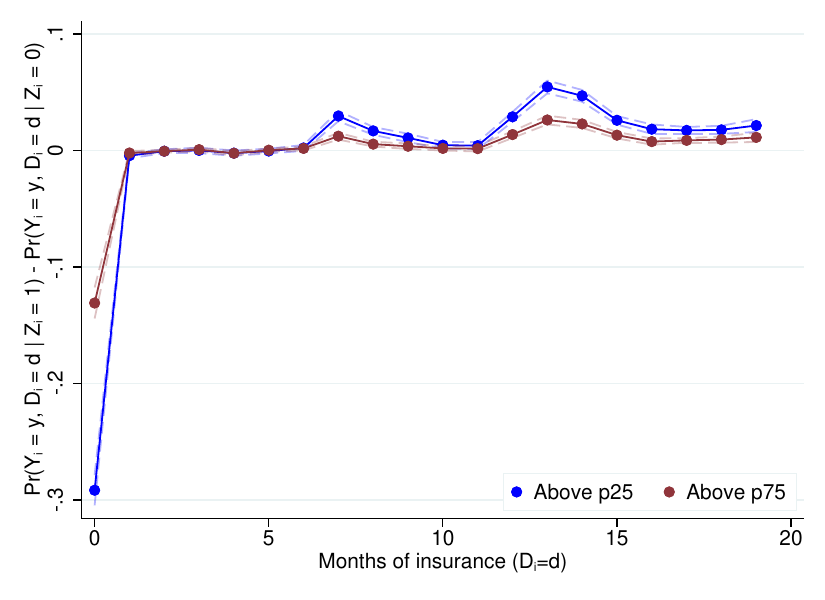}
    \\ 
    (a) Proposition \ref{prop:SDC_monotonicity} moment conditions  
    & 
    (b) Proposition \ref{prop:Ymoment_densityPositive} moment conditions 
    \end{tabular}
    \bigskip
    \subcaption*{ \emph{Notes:} \footnotesize This figure shows visual tests of Propositions \ref{prop:SDC_monotonicity} and  \ref{prop:Ymoment_densityPositive}. Panel (a) estimates the difference in density of months of insurance coverage between individuals lotteried into Medicaid ($Z_i=1$) and those not ($Z_i=0$). Consistent with EMCO, there is a large decrease in mass at $D_i=0$ and positive increases for all $D_i > 0$. Panel (b) estimates differences in the intersection of outcome values and treatment levels. The outcome is a binary indicator for whether a standardized measure of health care utilization falls above sample quantiles. Consistent with EMCO, there is a large decrease at $D_i=0$ and positive increases elsewhere. The figure was produced using the replication data from \cite{finkelstein2012}. Since administrative data on mortality and hospitalization are unavailable in the public data, we use the survey-sample and estimate differences by regressing indicators for $D_i=d$ or $(D_i=d)(Y_i =y)$ on the lottery dummy controlling for lottery strata, as in the original analysis. Standard errors are clustered at the household level.}
    \end{figure}
\else
%   False
\fi

 Tests of EMCO described in the Online Appendix support its applicability to the OHIE. Appendix Figure \ref{fig:ohie1}, for example, shows that individuals lotteried into Medicaid ($Z_i=1$) were much less likely than controls to have zero months of insurance over the follow-up period. They are more likely, however, to have coverage for all positive durations, with particularly large spikes around 6-7 months and 12-13 months. EMCO's restrictions on treatment choices require that there are no decreases in density at any positive level of coverage.\footnote{Formal tests of the restrictions described in the Online Appendix show that we cannot reject at the 5\% significance level the null hypothesis that the data are consistent with EMCO ($T_n=3.46$; critical value $3.96$).}

EMCO gives the \cite{finkelstein2012} results a simple causal interpretation. The two percentage point increase in hospital admissions due to Medicaid presented in their Table IV, for example, reflects increases relative to how often subjects would have been admitted without any insurance. EMCO also allows the researcher to further decompose treatment effects across complier populations. Table \ref{ohietable} illustrates this using survey data from \cite{finkelstein2012} on self-reported health and health care utilization. Since there are many survey questions relevant for these outcomes, we create standardized indices by averaging normalized answers to seven health questions and four utilization questions (as in Table V and IX in \cite{finkelstein2012}). Column 1 shows 2SLS estimates of the effect of any Medicaid on these indices. Consistent with the original results, Medicaid increases both utilization and health significantly. The former rises by 0.1 standard deviations, and the latter by 0.2.

Under EMCO, all compliers in the OHIE are shifted from zero months of Medicaid to some positive amount. Column 2 shows the average untreated outcomes for all compliers. Compliers appear to be significantly negatively-selected on health---untreated means are 0.15 standard deviations below the sample average. Utilization is slightly higher than the sample average, but not significantly so. Columns 3-6 then show treated mean outcomes for compliers shifted to 7-12 months of Medicaid, 13-16 months, etc. No individuals were shifted to 1-6 months of coverage. The final row of the table reports the share of all compliers each group comprises. Roughly 33\% of compliers, for example, are shifted from zero months to 7-12 months. 

These means several interesting patterns. For example, all complier groups with positive density have higher treated health than the average under no insurance. But compliers induced to remain on Medicaid for the longest have the worst treated health. This is consistent with the adverse selection patterns noted in \cite{finkelstein2012} and \cite{kowalski2021reconciling}---only the sickest remain on Medicaid for long.\footnote{For example, comparing OLS to 2SLS estimates, the authors note that ``differences suggest that at least within a low-income population, individuals who select health insurance coverage are in poorer health (and therefore demand more medical care) than those who are uninsured, just as standard adverse selection theory would predict."} Compliers who remain on Medicaid for 7-12 months, by contrast, have better than average treated health. The 2SLS estimate in Column 1 is the population share-weighted average of means in Columns 3-5 minus the average in Column 2. As noted above, means under no insurance for each d-type complier group are not identified. The data are consistent with any pattern of treatment effects where the weighted average of complier means under no insurance equals the means in Column 2. It is possible, for example, that treatment effects are negative for compliers induced to stay on Medicaid the longest. Any negative effects for this group, however, would have to be outweighed by positive effects for others.

Utilization follows a similar pattern. Individuals induced to remain on Medicaid for the longest also have the highest levels of treated utilization. Thus the sickest compliers are also the most expensive. Utilization results also suggest that increases in utilization due to Medicaid are not due to one-shot ``pent-up" demand, since utilization is increasing in duration on Medicaid. As with health outcomes, the data are consistent with a range of treatment effects on utilization for each complier group.

Taken together, the results show the power of EMCO in a setting where it is institutionally plausible. Not only do the simple 2SLS regressions presented in \cite{finkelstein2012} have a coherent causal interpretation, but identification of complier means reveal important insights into selection patterns.

\ifdefined\something
    \begin{table} 
    \caption{Decomposition of Treatment Effects in the OHIE}  \label{ohietable}
    \input{ohie_combined.tex}
    \bigskip
    \subcaption*{\emph{Notes:} \footnotesize This table shows 2SLS effects of any Medicaid and complier mean outcome levels for health and utilization indices. The health index is the average of seven survey questions on health, each normalized to be mean zero and standard deviation 1, and likewise with four questions on utilization. Positive values of each reflect improved mental and physical health and increased utilization, respectively. Column 1 shows the 2SLS estimate of the effect of any Medicaid on each index. As in \cite{finkelstein2012}, the specification includes lottery fixed effects and clusters standard errors at the household level. Column 2 shows complier mean outcomes under no insurance. They are estimated from an identical 2SLS specification with $Y_i1(D_i = 0)$ as the outcome and $1(D_i = 0)$ as the endogenous variable. Columns 3-6 show mean outcomes for compliers shifted from zero to the quantity of insurance listed in the header. They are estimated from 2SLS regressions with  $Y_i1(D_i = d)$ as the outcome and $1(D_i = d)$ as the endogenous variable. The final row of the table reports the share of each d-type compliers. For example, column 2 identifies the untreated mean for 100\% of compliers, since it averages across all d-types. Column 3 shows treated means for compliers shifted to 7-12 months, who comprise 33\% of all compliers.}
    \end{table}
\else
%   False
\fi

\FloatBarrier
\section{Conclusion} \label{sec:Conclusions}

2SLS estimates of the effects of ordered treatments (e.g., years of education) can be difficult to interpret. The estimand is a weighted average of causal effects for overlapping sets of compliers and may differ from the estimand of interest. In some settings, an auxiliary assumption that the instruments only induce units to switch from zero units of treatment to some positive amount (EMCO) may be reasonable. Under EMCO, 2SLS estimates of the effect of a recoded indicator for \textit{any} treatment capture average treatment effects for mutually exclusive compliers groups shifted into varying units of treatment. Treated means for each complier type can be recovered using standard techniques, as well as the average untreated mean, allowing for a partial decomposition of the estimand. An application to data from the Oregon Health Insurance Experiment reveals clear patterns of adverse selection into Medicaid. 

When should researchers invoke EMCO? While there are testable necessary implications of EMCO, in practice the assumption's validity is likely to hinge on the institutional details of the experiment at hand. Along with the standard LATE assumptions, invoking EMCO is equivalent to assuming the data is generated by a two-stage selection process where individuals first decide whether to participate in treatment at all and then pick treatment levels. EMCO requires that the instruments only affect the utility of any participation, but not the relative utility of positive treatment levels. In many settings, such as those with one-sided non-compliance only, it may be clear \textit{a priori} whether this is reasonable. When EMCO does hold, researchers should invoke it explicitly to justify their choice of models and make use of the additional identifying power it provides.

\clearpage
\singlespacing
\bibliographystyle{aer}
\bibliography{arxiv_mar2024}
\onehalfspacing

\FloatBarrier
\begin{table} 
\caption{Decomposition of Treatment Effects in the OHIE}  \label{ohietable}
\input{ohie_combined.tex}
\bigskip
\subcaption*{\emph{Notes:} \footnotesize This table shows 2SLS effects of any Medicaid and complier mean outcome levels for health and utilization indices. The health index is the average of seven survey questions on health, each normalized to be mean zero and standard deviation 1, and likewise with four questions on utilization. Positive values of each reflect improved mental and physical health and increased utilization, respectively. Column 1 shows the 2SLS estimate of the effect of any Medicaid on each index. As in \cite{finkelstein2012}, the specification includes lottery fixed effects and clusters standard errors at the household level. Column 2 shows complier mean outcomes under no insurance. They are estimated from an identical 2SLS specification with $Y1(D_i = 0)$ as the outcome and $1(D_i = 0)$ as the endogenous variable. Columns 3-6 show mean outcomes for compliers shifted from zero to the quantity of insurance listed in the header. They are estimated from 2SLS regressions with  $Y1(D_i = d)$ as the outcome and $1(D_i = d)$ as the endogenous variable. The final row of the table reports the share of each d-type compliers. For example, column 2 identifies the untreated mean for 100\% of compliers, since it averages across all d-types. Column 3 shows treated means for compliers shifted to 7-12 months, which comprises 33\% of all compliers. ***, **, and * indicate significance at the 1, 5 and 10 percent level, respectively.}
\end{table}

\FloatBarrier
\clearpage
\appendix
\counterwithin{figure}{section}
\counterwithin{table}{section}
\counterwithin{equation}{section}

\clearpage
\vspace*{\fill}
\begin{center}
    \section*{Appendix}
\end{center}
\vspace*{\fill}

\clearpage
\section{Proofs}
\label{appen:proofs}

\subsection{Proof of Proposition \ref{eq:abadi_betaJ_discrete}} \label{appen:proof_abadi_betaJ_discrete}

We begin by presenting a more general version of Proposition \ref{eq:abadi_betaJ_discrete} that is analogous to Theorem 3.1 in \cite{abadie2003semiparametric} and generalizes this result for the binary treatment case to multi-valued treatments under EMCO. A direct implication of Proposition \ref{prop:generalize_Abadi2003} is that compliers' observable pre-treatment characteristics can be identified as discussed in Equation \ref{eq:compliersX}.

\begin{proposition}{}\label{prop:generalize_Abadi2003} Let $g(\cdot)$ be a measurable function on $(Y_i,D_i,X_i)$ such that $\e{g(Y_i,D_i,X_i)} < \infty$. Assume that $\Pr(D_i(1) = d > D_i(0)) > 0$ $\forall \ d>0$ and that Assumptions \ref{ass:LATE_Assumptions} and \ref{ass:ExtensiveOnly} hold conditional on the covariates $X_i$. Then:

\begin{align*}
    (i) \; \frac{1}{\Pr(D_i(1) > D_i(0))} \e{\kappa_i g(Y_i,D_i,X_i)} &= \e{g(Y_i,D_i,X_i)|D_i(1)>D_i(0)=0}, \\
    (ii) \;  \frac{1}{\Pr(D_i(1) = d > D_i(0))} \e{\kappa_{i,(d)} g(Y_i,D_i,X_i)} &= \e{g(Y_i(d),d,X_i)|D_i(1)=d>D_i(0)=0} \; for \; d>0, \\
    (iii) \;  \frac{1}{\Pr(D_i(1) > D_i(0))} \e{\kappa_{i,(0)} g(Y_i,D_i,X_i)} &= \e{g(Y_i(0),0,X_i)|D_i(1)>D_i(0)=0}
\end{align*}
where: 
\begin{small}
\begin{align*}
\kappa_i &= 1 - \frac{1(D_i = 0) Z_i }{\Pr(Z_i=1 | X_i)} - \frac{ 1(D_i>0) \left( 1 - Z_i \right) }{\Pr(Z_i=0 | X_i)}
\\
\kappa_{i,(d)} &= 1(D_i = d) \frac{Z_i - \Pr(Z_i=1 | X_i)}{\Pr(Z_i=0 | X_i) \Pr(Z_i=1 | X_i)} , \quad for \quad d > 0 
\\
\kappa_{i,(0)} &= 1(D_i = 0) \frac{(1-Z_i) - \Pr(Z_i=0 | X_i)}{\Pr(Z_i=0 | X_i) \Pr(Z_i=1 | X_i)}
\end{align*}
\end{small}
\end{proposition}

\subsubsection*{Proof of part (i)}

Note that under the maintained assumptions, for any $s \subseteq \{0,\dots,\bar{D}\}$ and $z \in \{0,1\}$
\begin{flalign*}
\e{g(Y_i,D_i,X_i)1(D_i \in s)1(Z_i = z) | X_i} &= \e{g(Y_i,D_i,X_i) | D_i \in s, Z_i = z, X_i}\Pr(D_i \in s, Z_i=z|X_i) \\
&=\e{g(Y_i,D_i,X_i) | D_i(z) \in s, X_i}\Pr(D_i(z) \in s | X_i)\Pr(Z_i=z|X_i)
\end{flalign*}

\noindent Thus $\e{\kappa_i g(Y_i,D_i,X_i) | X_i}$ evaluates to
\begin{flalign*}
\e{\kappa_i g(Y_i,D_i,X_i) | X_i} &= \e{g(Y_i,D_i,X_i) | X_i} \\
&- \e{g(Y_i,D_i,X_i)|D_i(1) = 0,X_i} \Pr(D_i(1)=0|X_i) \\
&- \e{g(Y_i,D_i,X_i)|D_i(0) > 0,X_i} \Pr(D_i(0)>0|X_i) \\
\end{flalign*}

\noindent Iterating over compliance types possible under LATE and EMCO implies that
\begin{flalign}
    \label{eq:Abadi_0a}
    \e{g(Y_i,D_i,X_i) | X_i} &= \e{g(Y_i,D_i,X_i) 1(D_i(1) > D_i(0)) | X_i} \\
    &+ \e{g(Y_i, D_i,X_i)|D_i(1)= 0,X_i} \Pr(D_i(1)= 0 | X_i)\nonumber \\
    &+ \e{g(Y_i, D_i,X_i)|D_i(0) > 0,X_i} \Pr(D_i(0) > 0 | X_i)\nonumber 
    \end{flalign}

\noindent which implies that $$\e{\kappa_i g(Y_i,D_i,X_i) | X_i} = \e{g(Y_i,D_i,X_i)1(D_i(1) > D_i(0)) |X_i}$$

\noindent Applying the law of iterated expectations over $X_i$ shows that
\begin{flalign*}
\e{\kappa_i g(Y_i,D_i,X_i)} &= \e{\e{\kappa_i g(Y_i,D_i,X_i) | X_i} } \\
&=\e{g(Y_i,D_i,X_i)1(D_i(1) > D_i(0))} \\
&= \e{g(Y_i,D_i,X_i) | D_i(1) > D_i(0) = 0} \Pr(D_i(1) > D_i(0) = 0)
\end{flalign*}

\noindent Dividing by $\Pr(D_i(1) > D_i(0) = 0)$ completes the proof.

\subsubsection*{Proof of part (ii)}

By similar logic to that above, note that for any $d \in \{0,\dots,\bar{D}\}$ and $z \in \{0,1\}$:
\begin{flalign*}
&\e{g(Y_i,D_i,X_i)1(D_i = d)1(Z_i = z) | X_i} \\
&=\e{g(Y_i(d),d,X_i) | D_i(z) = d, X_i}\Pr(D_i(z)= d | X_i)\Pr(Z_i=z|X_i)
\end{flalign*}

\noindent Thus we have
\begin{align*}
     \e{ \kappa_{i,(d)}g(Y_i,D_i,X_i) | X_i} &=  \e{g(Y_i(d),d,X_i) | D_i(1) = d, X_i}\frac{\Pr(D_i(1) = d | X_i)}{\Pr(Z_i=0|X_i)} \\
      &- \e{g(Y_i(d),d,X_i) | D_i(1) = d, X_i}\frac{\Pr(D_i(1) = d | X_i)Pr(Z_i=1|X_i)}{\Pr(Z_i=0|X_i)} \\
      &- \e{g(Y_i(d),d,X_i) | D_i(0) = d, X_i}\Pr(D_i(0) = d | X_i) \\
      &=  \e{g(Y_i(d),d,X_i) | D_i(1) = d, X_i}\Pr(D_i(1) = d | X_i) \\
      &- \e{g(Y_i(d),d,X_i) | D_i(0) = d, X_i}\Pr(D_i(0) = d | X_i) 
\end{align*}

\noindent Under EMCO
\begin{align*}
    &\e{g(Y_i(d),d,X_i) | D_i(1) = d, X_i}\Pr(D_i(1) = d | X_i) \\ &= \e{g(Y_i(d),d,X_i) | D_i(1) = d > D_i(0), X_i}\Pr(D_i(1) = d > D_i(0) | X_i) \\
    &+ \e{g(Y_i(d),d,X_i) | D_i(0) = d, X_i}\Pr(D_i(0)  = d | X_i)
\end{align*}

\noindent Thus $ \e{ \kappa_{i,(d)}g(Y_i,D_i,X_i) | X_i}  = \e{g(Y_i(d),d,X_i) 1(D_i(1) = d > D_i(0) )| X_i}$. Applying iterated expectations as in the proof of part (i) completes the proof.

\subsubsection*{Proof of part (iii)}

The proof of part (iii) is similar. Following analogous steps to those in part (ii), we have:
\begin{align*}
     \e{ \kappa_{i,(0)}g(Y_i,D_i,X_i) | X_i} &=  \e{g(Y_i(0),0,X_i) | D_i(0) =0, X_i}\Pr(D_i(0) = 0 | X_i) \\
      &- \e{g(Y_i(0),0,X_i) | D_i(1) = 0, X_i}\Pr(D_i(1) = 0 | X_i) 
\end{align*}

\noindent EMCO implies that:
\begin{align*}
    &\e{g(Y_i(0),0,X_i) | D_i(0) = 0, X_i}\Pr(D_i(0) = 0 | X_i) \\ &= \e{g(Y_i(0),0,X_i) | D_i(1) > D_i(0) = 0, X_i}\Pr(D_i(1) > D_i(0) = 0 | X_i) \\
    &+ \e{g(Y_i(0),0,X_i) | D_i(1) = 0, X_i}\Pr(D_i(1) = 0 | X_i)
\end{align*}

\noindent Thus $ \e{ \kappa_{i,(0)}g(Y_i,D_i,X_i) | X_i}  = \e{g(Y_i(0),0,X_i) 1( D_i(1) > D_i(0) = 0 )| X_i}$. Applying iterated expectations as in the proof of part (i) completes the proof.

\subsection{Proof of Proposition \ref{prop:equivalenceHurdle}}
The proof that the model described in Assumption \ref{eq:hurdle_simple} satisfies LATE and EMCO is straightforward. Monotonicity, for example, is guaranteed by the structure of the threshold crossing model for treatment take-up $1( \pi_0(Z_i) - U_i^{Ext} \geq 0)$ and EMCO holds because $\pi_d$ does not depend on $Z_i$ for $d > 0$. Exogeneity and exclusion are assumed directly.

To show that LATE and EMCO imply the selection model representation in Assumption \ref{eq:hurdle_simple}, we will construct marginally uniform random variables $U_i^{Ext}$ and $U_i^{Int}$ that, when combined with Assumption \ref{eq:hurdle_simple}, satisfy all the assumptions of LATE and EMCO and imply the same set of observed and counterfactual treatment choices. This demonstrates that when EMCO holds, the LATE framework assumptions can be represented by a two-latent factor hurdle model without imposing any additional restrictions.

To begin, note that EMCO and LATE together restrict choice behavior to three categories of individuals: never-takers, who have $D_i(1)=D_i(0)=0$; $\bar{D}$ types of always-takers, each with $D_i(1)=D_i(0)=d$; and  $\bar{D}$ types of compliers, each with $D_i(1) = d > 0 = D_i(0)$. Let $U_i^{Ext}$ be a uniform random variable on $[0,1]$. Assuming without loss that $\pi_0(1) \leq \pi_0(0)$, assign the mass of each category of individuals to values of $U_i^{Ext}$ uniformly as follows:
\begin{enumerate}
\item Never-takers: $U_i^{Ext} \in [0,\pi_0(1)]$
\item Always-takers: $U_i^{Ext} \in (\pi_0(0),1]$,
\item Compliers: $U_i^{Ext} \in (\pi_0(1),\pi_0(0)]$
\end{enumerate}

where $\pi_0(1) = Pr(D_i = 0 | Z_i = 1) = Pr(D_i(1) = D_i(0) = 0), 1-\pi_0(0) = Pr(D_i > 0 | Z_i = 0) = \sum_{d=1}^{\bar{D}} Pr(D_i(1) = D_i(0) = d)$, and $\pi_0(0) - \pi_0(1) = Pr(D_i(1) > 0, D_i(0) = 0) = \sum_{d=1}^{\bar{D}} Pr(D_i(1) = d, D_i(0) = 0)$. The width of the supports for each category equal their population mass, so that the density of $U_i^{Ext}$ is one at each point in the support and ensuring $U_i^{Ext}$ is marginally uniform.

Likewise, let $U_i^{Int}$ be a uniform random variable on $[0,1]$ and assign each category of individuals values uniformly over the intervals:
\begin{enumerate}
\item Never-takers: $U_i^{int} \in [0, 1]$
\item Type-$d$ always-takers: $U_i^{int} \in [\pi_{d+1}, \pi_d)$
\item Type-$d$ compliers: $U_i^{int} \in [\pi_{d+1}, \pi_d)$
\end{enumerate}
where $\pi_1 = 1, \pi_{\bar{D}+1} = 0$, and $\pi_d - \pi_{d+1} = Pr(D_i = d | Z_i = 1, D_i > 0) = [Pr(D_i(1) = D_i(0) = d) + Pr(D_i(1) = d,  D_i(0) = 0)]/Pr(D_i(1) > 0)$. The width of the supports for each category of type-$d$ always-takers and compliers equals their combined population shares excluding never-takers. Mixing across the type-$d$ always-takers and compliers, the density of $U_i^{Int}$ will therefore be $1-Pr(D_i(1) = D_i(0) = 0) = Pr(D_i(1) > 0)$ across each $[\pi_{d+1},\pi_d)$ interval, which togeher span the unit interval. Adding the population of never-takers uniformly over $[0,1]$ ensures that $U_i^{Int}$ is marginally uniform as well.

Note that the uniform distributional assumptions are inconsequential, since for any strictly increasing set of distribution functions $G_1$ and $G_2$, we can construct a non-marginally uniform model with $\tilde{U_i}^{Ext} = G_1^{-1}(U_i^{Ext})$, $\tilde{\pi}_0(Z_i) = G_1^{-1}(\pi_0(Z_i))$, $\tilde{U_i}^{Int} = G_2^{-1}(U_i^{Int})$, and $\tilde{\pi}_d =  G_2^{-1}(\pi_d)$ for $d>0$. Thus is it only necessary that $U_i^{Ext}$ and $U_i^{Int}$ have strictly increasing marginal distribution functions, as stated in the proposition.

We have thus constructed a selection model that satisfies all the assumptions of LATE and EMCO and implies the same set of observed and counterfactual treatment choices. Adding the independence and exclusion assumption---$(Y_i(0),\dots,Y_i(\bar{D}),U_i^{Ext}, U_i^{Int}) \ \indep \ Z_i$---completes the equivalence. The LATE framework assumptions can therefore also be represented by a two-latent factor hurdle model.

\subsection{Proof of Proposition \ref{prop:singleIndex_and_EMCO}}

Suppose that $\pi_d(0) < \pi_d(1)$ for some $d > 0$. Then individuals with $U_i \in [\pi_d(0),\pi_d(1))$ have $D_i(0) < d$, $D_i(1) \geq d$, and therefore $D_i(1) > D_i(0)$. Assumption \ref{ass:ExtensiveOnly} requires that $D_i(1) > D_i(0) \rightarrow D_i(0) = 0$. Since  $1(D_i(0) = 0) = 1(\pi_{1}(0) \leq U_i < \pi_0(0))$, compatibility with EMCO requires that for individuals with $U_i \in [\pi_d(0),\pi_d(1))$, $\pi_{1}(0) \leq U_i$. Therefore $\pi_d(0) = \pi_1(0)$. Because $\pi_{d}(0) \leq \pi_{d-1}(0) \ \forall d > 0$, this can also be stated as $\pi_{d}(0) = \pi_{d'}(0) \ \forall \ d' \  s.t. \ 1 \leq d' < d$.

Compatibility with EMCO therefore requires that for every $d > 1$ either $\pi_d(0) = \pi_d(1)$ or $\pi_{d}(0) = \pi_{d'}(0) \ \forall \ d' \  s.t. \ 1 \leq d' < d$. The former implies that $Pr(D_i(0) \geq d) = Pr(D_i(1) \geq d)$. The latter implies that $Pr(D_i(0) \in \{1,\dots,d-1\}) =1-Pr(D_i(0) = 0) - Pr(D_i(0) \geq d) =  0$. Translating these requirements into population moments, single-index compatibility with EMCO therefore requires:  
$$\left(\e{1(D_i \geq d)|Z_i=1} - \e{1(D_i \geq d)|Z_i=0}\right)\left(1-\e{1(D_i \geq d)|Z_i=0} - \e{1(D_i = 0)|Z_i=0}\right) = 0$$ holds for all $d > 1$. Note that this condition holds mechanically for $d=0$ (the first parenthetical term is zero) and for $d=1$ (the second parenthetical term is zero). 

\appendix
\counterwithin{figure}{section}
\counterwithin{table}{section}
\counterwithin{equation}{section}

\clearpage
\setcounter{page}{1}
\vspace*{2cm}
\begin{center}
    \section*{Online Appendix}
\end{center}
\vspace*{1cm}

The online appendix includes the following sections: \\

\begin{itemize}
    \item \textbf{Appendix \ref{sec: additional_figures}}. This section includes additional figures referenced in the main text.
    \item \textbf{Appendix \ref{sec: testable}}. This section discusses tests of EMCO and presents Monte-Carlos simulations investigating their power.  
    \item \textbf{Appendix \ref{appen:CvariateAdjust}}. This section discusses how covariates can be incorporated when testing the moment inequalities described in Appendix \ref{sec: testable}.
    \item \textbf{Appendix \ref{appen:proof_of_appendix_propositions}}. This section presents proofs of propositions in this online appendix.
\end{itemize}

\vspace*{\fill}

\FloatBarrier
\clearpage
\section{Additional figures} \label{sec: additional_figures}

\begin{figure}[!ht]{}
\caption{Visual Tests of EMCO in \cite{finkelstein2012}'s Analysis of the OHIE}  
\label{fig:ohie1}
\begin{tabular}{cc}
\includegraphics[scale=0.56]{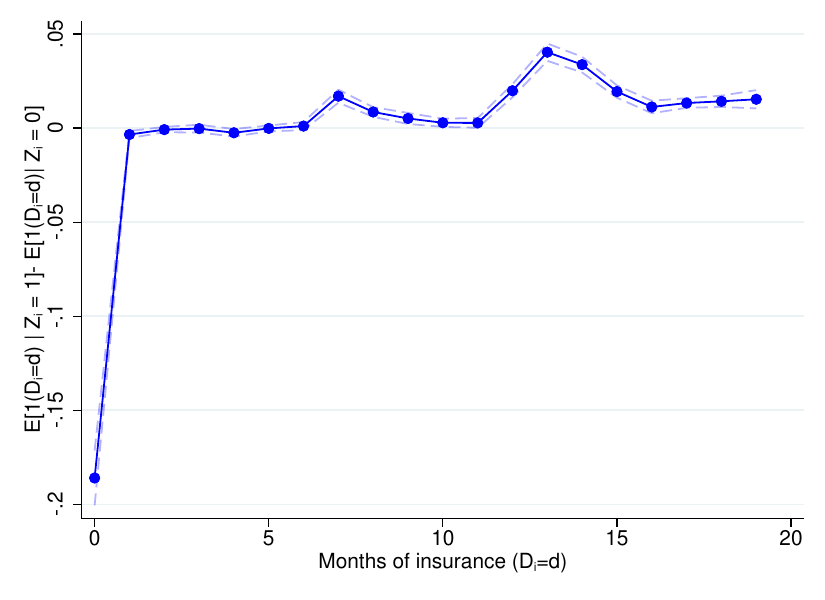}
&
\includegraphics[scale=0.56]{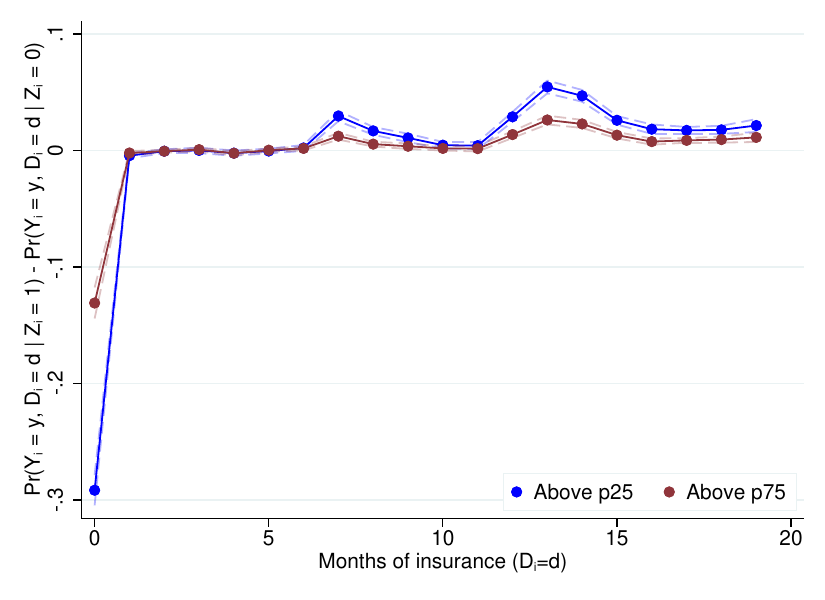}
\\ 
(a) Proposition \ref{prop:SDC_monotonicity} moment conditions  
& 
(b) Proposition \ref{prop:Ymoment_densityPositive} moment conditions 
\end{tabular}
\bigskip
\subcaption*{ \emph{Notes:} \footnotesize This figure shows visual tests of Propositions \ref{prop:SDC_monotonicity} and  \ref{prop:Ymoment_densityPositive} for the OHIE replication data \citep{finkelstein2012}. Panel (a) estimates the differences in the distribution of months of insurance coverage between individuals lotteried into Medicaid ($Z_i=1$) and those not ($Z_i=0$). Consistent with EMCO, there is a large decrease in mass at $D_i=0$ and positive increases for all $D_i > 0$. Panel (b) estimates differences in the intersection of outcome values and treatment levels. The outcome is a binary indicator for whether a standardized measure of health care utilization falls above sample quantiles. Consistent with EMCO, there is a large decrease at $D_i=0$ and positive increases elsewhere. Since administrative data on mortality and hospitalization are unavailable in the public replication data, we use the survey sample and estimate differences by regressing indicators for $D_i=d$ or $(D_i=d)(Y_i =y)$ on the lottery dummy controlling for lottery strata, as in the original analysis. Standard errors are clustered at the household level.}
\end{figure}

\FloatBarrier
\clearpage
\section{Testing EMCO} \label{sec: testable}

Combined with the LATE assumptions, EMCO places two sets of restrictions on the DGP. First, it restricts the distribution of treatments under $Z_i=1$ vs. $Z_i=0$. Specifically, EMCO implies that individuals are only shifted from no treatment into positive levels of treatment, ruling out any movement of individuals already at a positive level of treatment. Thus, there cannot be a \textit{decrease} in observations at positive treatment levels due to $Z_i$:
\begin{proposition}{}
\label{prop:SDC_monotonicity}
If Assumptions \ref{ass:LATE_Assumptions} and \ref{ass:ExtensiveOnly} hold, then: \\ 
(i) $\Pr(D_i=0|Z_i=1) < \Pr(D_i=0|Z_i=0)$ \\
(ii) $\Pr(D_i=d|Z_i=1) \geq \Pr(D_i=d|Z_i=0)  \quad \forall d > 0$
\end{proposition}

Part (i) of Proposition \ref{prop:SDC_monotonicity} follows because Assumption \ref{ass:LATE_Assumptions} part (iii) (monotonicity) implies the instrument cannot increase treatment at its lowest level ($d=0$). Assumption \ref{ass:LATE_Assumptions} part (i) (relevance) combined with Assumption \ref{ass:ExtensiveOnly} implies that the instruments must decrease treatment at $d=0$, yielding the strict inequality in part (i) of Proposition \ref{prop:SDC_monotonicity}. Part (ii) of Proposition \ref{prop:SDC_monotonicity} is implied by the addition of Assumption \ref{ass:ExtensiveOnly}, but provides necessary and not sufficient conditions for it to hold.

To understand why, it is useful to consider a simple example. Table \ref{tab:simplest_SDC_example} presents a DGP for an ordered treatment with three levels. There are three potential types of compliers: Those shifted from $D_i(0)=0$ to $D_i(1)=1$ (with population share $\Delta_{ext}^1$), those shifted from $D_i(0)=0$ to $D_i(1)=2$ (share $\Delta_{ext}^2$), and those shifted from $D_i(0)=1$ to $D_i(0)=2$ (share $\Delta_{int}$). The existence of a positive quantity of this final complier type would violate EMCO. When $\Delta_{int}=0$ and $\Delta_{ext}^1, \Delta_{ext}^2 \geq 0$, the distribution of $D_i$ conditional on $Z_i=1$ stochastically dominates the distribution of $D_i$ conditional on $Z_i=0$ for all $d>0$. If $\Delta_{int} > \Delta_{ext}^1$, condition (ii) will be violated and EMCO will be rejected. But if $\Delta_{ext}^1 \geq \Delta_{int}>0$, condition (ii) will be satisfied even though EMCO is violated.

\begin{table}[!ht]
\caption{Illustration of Proposition \ref{prop:SDC_monotonicity}}
\label{tab:simplest_SDC_example}
\begin{tabular}{ccc}
Treatment & $\Pr(D_i(0)=d)$ & $\Pr(D_i(1)=d)$\\
($d$) & & \\
\hline
0 & $a$ & $a - \Delta_{ext}^1 - \Delta_{ext}^2$ \\
1 & $b$ & $b + \Delta_{ext}^1 - \Delta_{int}$ \\
2 & $1 - a - b$ & $1 -a - b + \Delta_{ext}^2 + \Delta_{int}$ \\
\hline \hline
\end{tabular} 
\end{table}

Proposition \ref{prop:SDC_monotonicity} suggests a very simple visual test of EMCO. When plotting the distribution of treatments under $Z_i=1$ and $Z_i=0$, the former should stochastically dominate for $d >0$. In research designs where the instrument is only conditionally randomly assigned, the researcher can plot estimates of the effect of $Z_i$ on $1(D_i \geq d)$ conditional on a specific set of controls and examine whether the coefficients are uniformly non-negative.\footnote{\cite{eckhoff2018instrument} consider a treatment defined by $1\{D_i \geq d^*\}$ and assume that the instrument only shifts individuals from below $d^*$ to above it (i.e., $D_i(0) < D_i(1) \Rightarrow D_i(0) < d^* \leq D_i(1)$. They show this assumption implies $\Pr(D_i(1) \geq d' > D_i(0)) \leq \Pr(D_i(1) \geq d > D_i(0))$ for every $d'>d>d^*$. The same result was also independently derived in \cite{rose_shemtov2019WIP}, Footnote 8, for the case where $d^* = 0$. As we show in Appendix \ref{sec: propsixequiv}, these conditions are equivalent to those in Proposition \ref{prop:SDC_monotonicity} part (ii). \cite{eckhoff2018instrument} also state that these conditions should apply conditional on each value of the outcome.}

EMCO also places a set of restrictions on the joint distribution of $(Y_i,D_i,Z_i)$. Specifically, the same arguments made in \cite{balkepearl97} and \cite{heckman2005} for the binary treatment case imply that:
\begin{proposition}{}
\label{prop:Ymoment_densityPositive}
If Assumptions \ref{ass:LATE_Assumptions} and \ref{ass:ExtensiveOnly} hold, then for any measurable set $A$ in the support of $Y_i$: \\ 
(i) $\Pr(Y_i \in A, D_i=0|Z_i=0) \geq \Pr(Y_i \in A, D_i=0|Z_i=1)$ \\
(ii) $ \Pr(Y_i \in A, D_i=d|Z_i=1) \geq \Pr(Y_i \in A, D_i=d|Z_i=0) \quad \forall d > 0$ 
\end{proposition}

Intuitively, because EMCO requires that the instrument induces shifts from $D_i=0$ to $D_i > 0$ only, the density of $Y_i$ across its support must be weakly increasing due to $Z_i$ for all individuals with $D_i > 0$, and decreasing due to $Z_i$ for those with $D_i = 0$. Proposition \ref{prop:SDC_monotonicity} and \ref{prop:Ymoment_densityPositive} are closely related because Proposition \ref{prop:SDC_monotonicity} implies Proposition \ref{prop:Ymoment_densityPositive} if $A$ is set to $\mathcal{Y}$. Because the inequality in part (i) of Proposition \ref{prop:Ymoment_densityPositive} is weak however, it is implied by the exogeneity and monotonicity components of Assumption \ref{ass:LATE_Assumptions} alone. Part (ii) is implied by the addition of Assumption \ref{ass:ExtensiveOnly} to Assumption \ref{ass:LATE_Assumptions}.

Proposition \ref{prop:Ymoment_densityPositive} can also be tested visually by comparing densities of $Y_i$ conditional on $D_i=d$ for $Z_i=1$ and $Z_i=0$. Just as with Proposition \ref{prop:SDC_monotonicity}, however, Proposition \ref{prop:Ymoment_densityPositive} is necessary but not sufficient for EMCO. To see why, note that:
\begin{align*}
\e{ 1(Y_i \in A)1(D_i=d) | Z_i=1}-\e{1(Y_i \in A) 1(D_i=d) | Z_i=0} &= \\
\Pr(Y_i(d) \in A , D_i(1) = d > D_i(0) = 0) & \ + \\
\Pr(Y_i(d) \in A , D_i(1) = d > D_i(0) > 0) &- \Pr(Y_i(d) \in A , D_i(1) > D_i(0) = d)
\end{align*}

Under EMCO, the final two terms on the right-hand side disappear; the difference on the left-hand side must be non-negative, since the remaining term on the right-hand side is a density. When EMCO does not hold, however, this quantity can still be positive whenever $\Pr(Y_i(d) \in A , D_i(1) > d = D_i(0))$ is not too large. Hence there are some DGPs under which EMCO fails but both propositions are satisfied. The ability to detect violations of EMCO depends on both the share of compliers with $D_i(0) > 0$ and the density of such individuals in regions of $Y_i$. 

One can test the restrictions in Propositions \ref{prop:SDC_monotonicity} and \ref{prop:Ymoment_densityPositive} by asking whether they appear consistent with the data. In particular, Proposition \ref{prop:Ymoment_densityPositive} implies that for any measurable $A \subseteq \mathcal{Y}$:
\begin{subequations}
\begin{flalign} 
\e{ 1(D_i = 0) \cdot 1(Y_i\in A) | Z_i = 1} - \e{1(D_i = 0) \cdot 1(Y_i \in A) | Z_i = 0} & \leq 0 
\label{eq:moments_all_mean:c} \\
\e{1(D_i = d) \cdot 1(Y_i \in A) | Z_i = 0} - \e{ 1(D_i = d) \cdot 1(Y_i \in A) | Z_i = 1} & \leq 0 \quad \forall \; d > 0 
\label{eq:moments_all_mean:d} 
\end{flalign}
\end{subequations}

For each $A$, Equations \ref{eq:moments_all_mean:c} and \ref{eq:moments_all_mean:d} imply $\bar{D}+1$ total restrictions. These restrictions must hold for any $A$, including $A = \mathcal{Y}$. If $Y_i$ is discrete, they must hold at each point in support of $Y_i$, yielding $(\bar{D} + 1) \cdot | \mathcal{Y} |$ inequalities. Adding additional subsets of $\mathcal{Y}$ (including $A = \mathcal{Y}$) adds only redundant inequalities that can be obtained by summing \ref{eq:moments_all_mean:c} and \ref{eq:moments_all_mean:d} over points in the support of $Y_i$. The restrictions in Equations \ref{eq:moments_all_mean:c} and \ref{eq:moments_all_mean:d} must also hold for any outcome. It is also possible to incorporate covariates, as discussed below in Appendix \ref{appen:CvariateAdjust}. Doing either can potentially yield more powerful tests.\footnote{These restrictions also capture a direct implication of Assumption \ref{ass:LATE_Assumptions} shown by \cite{angrist1995} to hold regardless of whether or not EMCO is true: $\e{1(D_i \geq d) | Z_i = 1} - \e{1(D_i \geq d) | Z_i = 0} \geq 0$ for all $d$. This can be seen by taking $A = \mathcal{Y}$ and summing Equations \ref{eq:moments_all:c} and \ref{eq:moments_all:d} appropriately across $d$.}

Testing the restrictions in Equations \ref{eq:moments_all:c}-\ref{eq:moments_all:d} implies testing the combined null hypothesis that \emph{both} the LATE and EMCO assumptions are satisfied. The tests described below can thus be thought of as an omnibus test for the joint null hypothesis that Assumption \ref{ass:LATE_Assumptions} and Assumption \ref{ass:ExtensiveOnly} hold. Rejection can indicate violations of either assumption or both.

We consider two approaches to testing these restrictions. The first approach is based on the observation that for a set of choices of subsets $A$, the restrictions in Equations \ref{eq:moments_all_mean:c}-\ref{eq:moments_all_mean:d} define a set of $p$ differences in means comprised of the $\bar{D}+1$ differences for each $A$. We denote the sample analogs of these differences $\bar{\delta} = (\bar{\delta}_1, \bar{\delta}_2,\dots, \bar{\delta}_p)$. When $p$ is fixed and the sample is sufficiently large, it is reasonable to invoke the asymptotic approximation that $\bar{\delta} \sim N(\delta, \Sigma)$. A simple but conservative test of the inequality restrictions in Equations \ref{eq:moments_all_mean:c}-\ref{eq:moments_all_mean:d} compares the maximum of the $p$ elements of $\bar{\delta}$ to the distribution of the largest element from a $N(0,\Sigma)$ random vector. We implement this approach in our simulations below, using the nonparametric bootstrap to estimate $\Sigma$. 

The second approach leverages tools from the moment inequality literature. If there are many outcomes, the outcomes are continuous or take on many values, or if the researcher wishes to incorporate a large set of covariates, the resulting set of moments can be large and it may be more prudent to use tools that are better suited to testing many inequality restrictions simultaneously. To see how this can be done, define for any measurable $A \subseteq \mathcal{Y}$:
\begin{subequations}
\begin{align} 
m_{i0} &= \frac{1(D_i = 0) \cdot 1(Y_i\in A) \cdot Z_i}{\Pr(Z_i=1)}  - \frac{1(D_i = 0) \cdot 1(Y_i \in A) \cdot (1-Z_i)}{1-\Pr(Z_i=1)}  
\label{eq:moments_all:c} \\
m_{id} &= \frac{1(D_i = d) \cdot 1(Y_i \in A) \cdot (1-Z_i) }{1-\Pr(Z_i=1)} 
- \frac{1(D_i = d) \cdot 1(Y_i \in A) \cdot Z_i}{\Pr(Z_i=1)}, \quad \text{for} \ \ d > 0
\label{eq:moments_all:d} 
\end{align}
\end{subequations}

Proposition \ref{prop:Ymoment_densityPositive} requires that: 
\begin{align*}
    \e{m_{id}} \leq 0 \quad \text{for} \quad d=0,\dots,\bar{D}.
\end{align*} 
The tools we use to test this hypothesis require the sequence of random vectors $m_i = \{m_{i0},\dots,m_{i{\bar{D}}}\}, i=1,\dots,N$ to be independently distributed, which complicates replacing $\Pr(Z_i=1)$ with its sample analog. For this test, we therefore proceed treating $\Pr(Z_i=1)$ as known (as would be the case, for example, in an experiment where the researcher controls assignment of the instrument, or as may be a reasonable approximation when the sample is large). Because $E[m_i] \leq 0$ for any measurable $A \subseteq \mathcal{Y}$, one can concatenate these random vectors for multiple choices of $A$, yielding a total of $p$ restrictions.

The econometrics literature has developed a variety of methods for testing moment inequalities of this form \citep[see][for a recent review]{canay2017practical}. We use a Kolmogorov-Smirnov (KS) variance adjusted test statistic given by:
\begin{align*}
T_n = \max \left\{ \underset{1 \leq j \leq p}{\max} \left\{\frac{\sqrt{N} \bar{m}_j }{S_j} \right\}, 0 \right\}
\end{align*}
where $j$ indexes the elements of $m_i$, $\bar{m}_j$ is the sample average of $m_{ij}$, and  $S^2_j = \frac{\sum_{i=1}^N (m_{ij} - \bar{m}_j)^2}{N-1}$ is an estimate of its variance.
The KS test statistic is commonly used in the literature on testing moment inequalities \citep[e.g.,][]{bai2019practical}. It is powerful when the objective is to detect whether any of the inequalities are violated \citep{CCK2018}, as is the case in our setting.

To obtain critical values, we use two recently proposed procedures that are computationally tractable and work well in settings where the number of moments is potentially larger than the number of observations. The first is \cite{RSW2014}'s two-step moment recentering approach. The second is \cite{CCK2018}, who proposed a two-step bootstrap procedure based on moment selection \citep{andrews2010inference,andrews2012inference,andrews2013inference}.\footnote{\cite{CCK2018} also propose methods based on self-normalized sums that allow one to analytically calculate critical values and are faster computationally. However, these approaches are generally less powerful than the bootstrap procedure. We use the bootstrap procedure (as in \cite{bai2019practical}) when comparing the power of the test proposed by \cite{CCK2018} to that proposed by \cite{RSW2014}.}  
Both of the tests require two steps. In the first step, a moment recentering or selection procedure takes place. The second step conducts inference on the test statistic using the bootstrap over the recentered or selected moments.\footnote{Other methods proposed recently for inference on and testing of moment inequality models include \cite{andrews2019inference}'s for linear conditional moment inequalities, \cite{chernozhukov2015constrained}, and \cite{cox2020simple}.}

\subsection{Simulation evidence} \label{sec: sim_evidence}

To explore the power of these tests, we construct a simple simulation based on the three treatment example in Table \ref{tab:simplest_SDC_example}. We assume the instrument $Z_i$ is binary and that there is a binary outcome $Y_i$.

Compliance types are as follows:
\begin{enumerate}
    \item Non-compliers ($D_i(1)=D_i(0)$), with population share $P_{non-compliers}$. 
    \item Type 1 extensive margin compliers ($D_i(1)=1>D_i(0)=0)$), with population share $\Delta_{ext1}$.
    \item Type 2 extensive margin compliers ($D_i(1)=2>D_i(0)=0)$), with population share $\Delta_{ext2}$.
    \item Intensive margin compliers ($D_i(1)=2>D_i(0)=1)$), with population share $\Delta_{int}$. 
\end{enumerate}

The distribution of treatment under $Z_i=1$ and $Z_i=0$ is deterministic for each type of complier. Non-compliers, whose treatment does not depend on $Z_i$, are assigned to each level of $D_i$ with equal probability. The remainder of the DGP is:
\begin{itemize}
    \item $Z_i \sim Bernoulli(0.5)$.
    \item $Y_i \sim Bernoulli(0.3)$ for all individuals who are not intensive margin compliers.
    \item $Y_i \sim Bernoulli(0.3 + \Delta_y)$ for intensive margin compliers.
\end{itemize}

We simulate from this DGP 1,000 times for several different values of $\Delta_{int}$ and $\Delta_y$ using 1,000 observations in each simulation. Each simulation holds $P_{non-compliers}$ fixed at 0.4. For simplicity, we also fix the probability of being assigned to each of the extensive margin complier types to be the same, so that $\Delta_{ext1}=\Delta_{ext2}=\Delta_{ext}$. Thus the likelihood of being an extensive margin complier of either type can be expressed as $\Delta_{ext}= (1 - \Delta_{int} - 0.4)/2$. In each iteration, we draw individual types from a multinomial distribution defined by their population shares.

To formally test EMCO's restrictions in each simulation, we use the asymptotic approximation to the distribution of the sample mean differences discussed above, and the moment inequality methods proposed by \cite{CCK2018} and \cite{RSW2014}. Since $Y_i$ is discrete, we evaluate the restrictions for $A = \{0\}$ and $A=\{1\}$.  Bootstraps use 1,000 iterations in each simulation. As discussed above, the moment inequality tests treat $\Pr(Z_i=1)$ as known, although replacing it with its sample analog happens to change simulation results little in this particular case.

\begin{figure}[!ht]
\caption{Simulations of tests of Assumptions \ref{ass:LATE_Assumptions} and \ref{ass:ExtensiveOnly}}  
\label{fig:sims_Simplest}
\begin{adjustbox}{center,width=\textwidth}
\begin{tabular}{cc}
\includegraphics[scale=0.56]{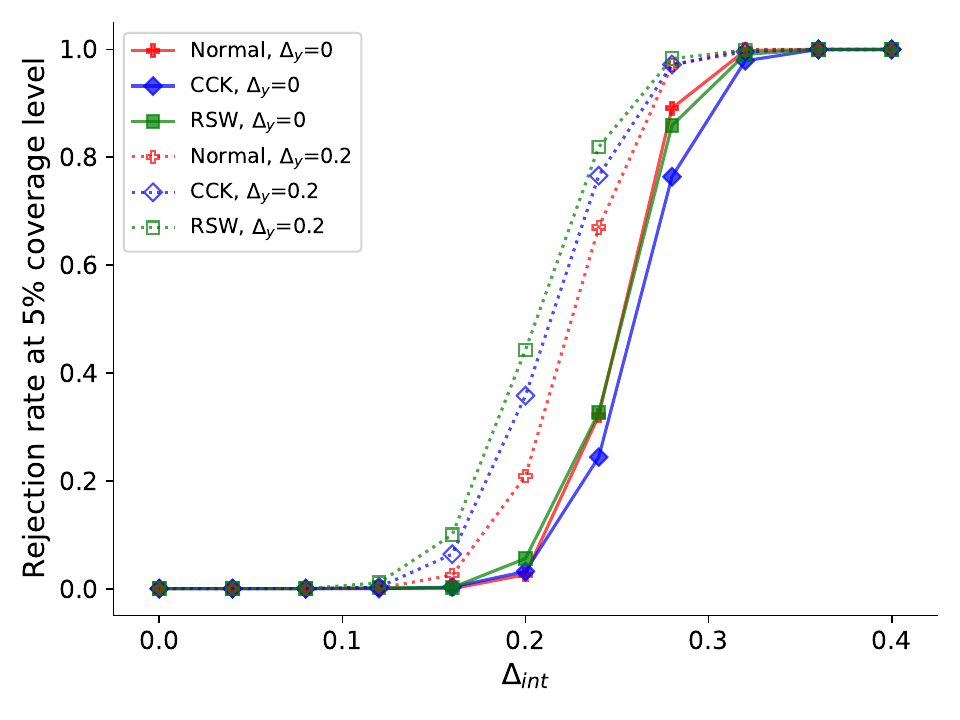}
&
\includegraphics[scale=0.56]{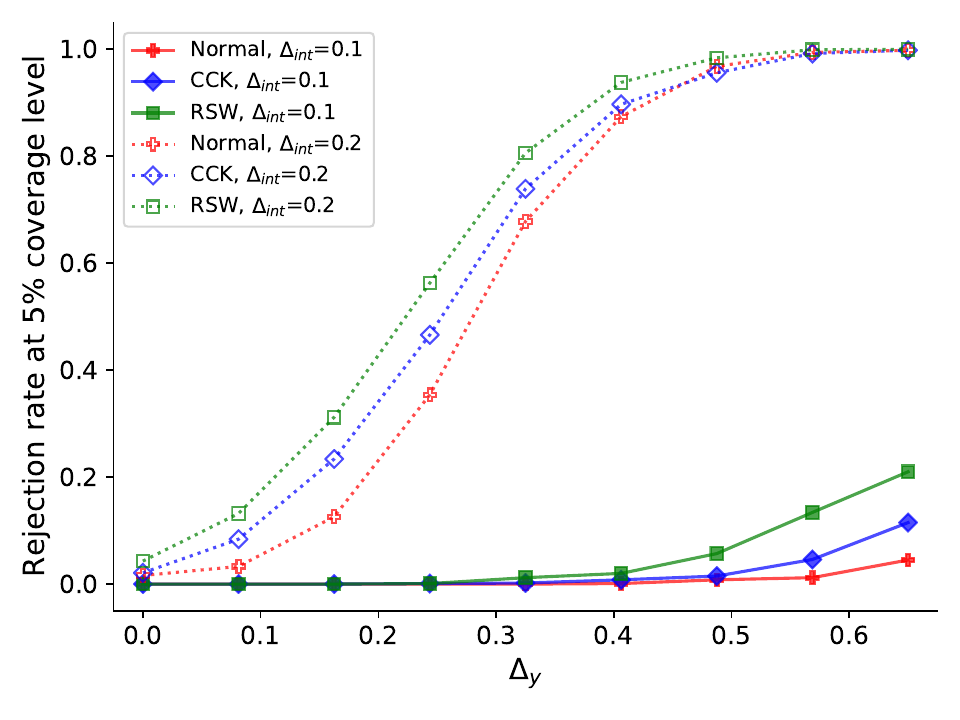} 
\\ 
(a) Varying the share of  
&
(b) Varying the difference in   
\\
intensive margin  compliers ($\Delta_{int}$) & outcome distributions ($\Delta_y$)
\end{tabular}
\end{adjustbox}
\bigskip
\subcaption*{ \emph{Notes:} \footnotesize This figure reports results from Monte-Carlo simulations based on the data generating process (DGP) described in Section \ref{sec: sim_evidence}.
The results average across 1,000 simulations for each combination of parameter values, with 1,000 observations used in each simulation. Each simulation holds the population share of compliers of all types fixed at 0.6. Panel (a) varies the proportion of intensive margin compliers ($\Delta_{int}$), with extensive margin compliers comprising the remainder of all compliers. The y-axis shows the share of simulations in which the null hypothesis that $\Delta_{int} = 0$ is rejected. The solid series plots results from simulations where all compliers have the same outcome distribution, while the dotted lines plot cases where $\Pr(Y_i=1)$ for intensive margin compliers differs by $\Delta_y$. Panel (b) plots results from the analogous exercise varying $\Delta_y$ while holding $\Delta_{int}$ fixed at either 10 or 20 percent.
``Normal" indicates the testing procedure described in Section \ref{sec: testable} that compares the maximum sample moment to its estimated distribution assuming all moment inequalities bind with equality. RSW indicates results using \cite{RSW2014}'s moment inequality testing procedure. CCK indicates \cite{CCK2018}'s method. These latter two procedures treat $\Pr(Z_i=1)$ as known.
}
\end{figure}

Figure \ref{fig:sims_Simplest} reports the results. In Panel (a), the x-axis shows the true proportion of intensive margin compliers ($\Delta_{int}$). The y-axis shows the share of simulations in which the null hypothesis that $\Delta_{int} = 0$ is rejected. One series holds $\Delta_y$ fixed at zero, while the other considers $\Delta_y = 0.2$. 
Panel (b) holds fixed the proportion of intensive margin compliers at either 10 or 20 percent and examines how power increases with differences in the distribution of outcomes, $\Delta_{y}$.  

The figure shows that both tests have good finite-sample power to detect  violations of EMCO large enough that the conditions in either Proposition \ref{prop:SDC_monotonicity} or \ref{prop:Ymoment_densityPositive} are violated. In these simulations, part (i) of Proposition \ref{prop:SDC_monotonicity} and part (ii) for $d=2$ always hold in the population. Varying $\Delta_{int}$ only affects part (ii) for $d=1$, which will be violated in this case if $\Delta_{int} > 0.2$. As result, power is effectively zero when $\Delta_{int}$ is below $0.2$ and $\Delta_y = 0$, as shown in Panel (a). Rejection rates increase sharply as $\Delta_{int}$ increases beyond $0.2$, however, and EMCO is almost always rejected when at least 30\% of individuals are intensive margin compliers. Non-zero $\Delta_y$ yields more power because it can generate violations of part (ii) of Proposition  \ref{prop:Ymoment_densityPositive} for $d=1$ even if the conditions in Proposition  \ref{prop:SDC_monotonicity} are satisfied.\footnote{Part (i) of Proposition \ref{prop:Ymoment_densityPositive} always holds in this DPG. So does part (ii) for $d=2$.} When $\Delta_y = .2$, for example, the condition is violated for $A = \{1\}$ and $d=1$ if $\Delta_{int} > .14$. All three  procedures preform similarly, although as expected the test based on asymptotic normality tends to be slightly more conservative.

Panel (b) further demonstrates how holding fixed $\Delta_{int}$, increases in $\Delta_y$ can increase rejection rates. When $\Delta_{int} = .1$, for example, the conditions in Proposition \ref{prop:SDC_monotonicity} are not violated, as discussed above. But if $\Delta_y$ is sufficiently large (in this case greater than about $.45$), the conditions in Proposition \ref{prop:Ymoment_densityPositive} kick in and EMCO can be rejected. Thus, both differences in the shares of intensive and extensive margin compliers and differences in their distribution of outcomes can be used to detect violations of EMCO. EMCO is most likely to be rejected when the share of intensive-margin compliers is large and their outcome distribution differs strongly from that of extensive-margin compliers.

\FloatBarrier
\clearpage
\section{Including covariates in tests of EMCO} \label{appen:CvariateAdjust}

Incorporating covariate adjustment into the moment inequality tests (and visualizations) is simple to implement. One can interact the random variables in Equations \ref{eq:moments_all:c}-\ref{eq:moments_all:d} with specific values of the covariates $X_i=x$. This can be used to construct additional and even more powerful tests of EMCO. For example, the moment condition defined by the expectation of Equation \ref{eq:moments_all:c} must hold for every $X_i=x$:

\begin{footnotesize}
\begin{align*}
 \e{  \frac{1}{\Pr(Z_i=1)} 1(D_i =0)  1(Y_i \in A) 1(X_i=x) Z_i - \frac{1}{1-\Pr(Z_i=1)} 1(D_i=0) 1(Y_i \in A) 1(X_i=x) (1-Z_i) }  & \leq 0
\end{align*}
\end{footnotesize}

Interacting the random variables in Equations \ref{eq:moments_all:c}-\ref{eq:moments_all:d} with different pre-treatment covariates will increase the number of moment inequalities. However, this is not necessarily a problem, since one of the advantages of the testing procedures proposed by \cite{RSW2014} and \cite{CCK2018} is that they can be used also in high-dimensional settings when the number of moment inequalities is potentially larger than the number of observations \citep{bai2019practical}. 

\FloatBarrier
\clearpage
\section{Proofs of propositions and claims in the online appendix} \label{appen:proof_of_appendix_propositions}

\subsection{Proofs of Propositions \ref{prop:SDC_monotonicity} and \ref{prop:Ymoment_densityPositive}}

To prove part (i) of Proposition \ref{prop:Ymoment_densityPositive}, note that under Assumptions \ref{ass:LATE_Assumptions} (LATE) and \ref{ass:ExtensiveOnly} (EMCO):
\begin{align*}
Pr(Y_i \in A, D_i = 0 | Z_i = 0) &= \Pr(D_i(0)=0, Y_i(0) \in A|Z_i=0) 
    \\ &= 
    \Pr(D_i(1)>0, D_i(0) = 0, Y_i(0) \in A)
    \nonumber \\ & \quad + \Pr(D_i(1)=0, D_i(0) = 0, Y_i(0) \in A)
    \nonumber \\
Pr(Y_i \in A, D_i = 0 | Z_i = 1) &= \Pr(D_i(1)=0, Y_i(0) \in A|Z_i=1) 
    \\ &= \Pr(D_i(1)=0, D_i(0) = 0, Y_i(0) \in A)
    \nonumber
\end{align*}

\noindent Thus $Pr(Y_i \in A, D_i = 0 | Z_i = 0) \geq Pr(Y_i \in A, D_i = 0 | Z_i = 1)$. Setting $A = \mathcal{Y}$ yields part (i) of Proposition \ref{prop:SDC_monotonicity} with a weak instead of strict inequality. Relevancy (Assumption \ref{ass:LATE_Assumptions}, part (i)) requires that $\Pr(D_i(1)>0, D_i(0) = 0, Y_i(0) \in A) > 0$ for $A = \mathcal{Y}$, which shows why the inequality in part (i) of Proposition \ref{prop:SDC_monotonicity} must be strict.

The proof of part (ii) is analogous, since under Assumptions \ref{ass:LATE_Assumptions} (LATE) and \ref{ass:ExtensiveOnly} (EMCO) for $d > 0$:
\begin{align*}
Pr(Y_i \in A, D_i = d | Z_i = 0) &= \Pr(D_i(0)=d, Y_i(d) \in A|Z_i=0) 
    \\ &= 
    \Pr(D_i(1)=d, D_i(0) = d, Y_i(d) \in A)
    \nonumber \\
Pr(Y_i \in A, D_i = d | Z_i = 1) &= \Pr(D_i(1)=d, Y_i(d) \in A|Z_i=1) 
    \\ &= \Pr(D_i(1)=d, D_i(0) = d, Y_i(d) \in A) \\
    & \quad + \Pr(D_i(1)=d, D_i(0) = 0, Y_i(d) \in A)
    \nonumber
\end{align*}

Thus $Pr(Y_i \in A, D_i = d | Z_i = 1) \geq Pr(Y_i \in A, D_i = d | Z_i = 0)$ for $d > 0$. Part (ii) of Proposition \ref{prop:SDC_monotonicity} follows when $A = \mathcal{Y}$.

\subsection{Equivalence between conditions in Proposition \ref{prop:SDC_monotonicity} part (ii) and Anderson and Huber (2021) conditions} \label{sec: propsixequiv}

Suppose that  $\Pr(D_i = d|Z_i=1) \geq \Pr(D_i = d|Z_i=0) \ \forall d > 0$. LATE and ECMO implies that: 
\begin{align*}
    \Pr(D_i = d|Z_i=1) - \Pr(D_i = d|Z_i=0) &= \Pr(D_i(1)=d>D_i(0)=0) \geq 0 \ \forall d > 0
\end{align*}

\noindent For any $d'>d > 0$:
\begin{align*}
&\Pr(D_i(1) \geq d' > D_i(0)) - \Pr(D_i(1) \geq d > D_i(0)) \\
&=\sum_{k=d'}^{\bar{D}} \Pr(D_i(1)=k>D_i(0)=0) -  \sum_{l=d}^{\bar{D}} \Pr(D_i(1)=l>D_i(0)=0)  \\
&=\sum_{k=d}^{d'-1} \Pr(D_i(1)=k>D_i(0)=0) \geq 0
\end{align*}

\noindent Likewise, if $\Pr(D_i(1) \geq d' > D_i(0)) - \Pr(D_i(1) \geq d > D_i(0))$ for any $d'>d > 0$, then clearly: 
\begin{align*}
&\Pr(D_i(1) \geq d+1 > D_i(0)) - \Pr(D_i(1) \geq d > D_i(0)) \\
&=\sum_{k=d+1}^{\bar{D}} \Pr(D_i(1)=k>D_i(0)=0) -  \sum_{l=d}^{\bar{D}} \Pr(D_i(1)=l>D_i(0)=0)  \\
&=\Pr(D_i(1)=d>D_i(0)=0) = \Pr(D_i = d|Z_i=1) - \Pr(D_i = d|Z_i=0) \geq 0
\end{align*}
for any $d > 0$. Thus the conditions in part (ii) of Proposition 6 are equivalent to the claim that $\Pr(D_i(1) \geq d' > D_i(0)) - \Pr(D_i(1) \geq d > D_i(0))$ for any $d'>d > 0$.

\end{document}

%% file: ohie_combined.tex
{
\def\sym#1{\ifmmode^{#1}\else\(^{#1}\)\fi}
\begin{tabular}{l*{5}{c}}
\hline\hline
            &\multicolumn{1}{c}{(1)}         &\multicolumn{1}{c}{(2)}         &\multicolumn{1}{c}{(3)}         &\multicolumn{1}{c}{(4)}         &\multicolumn{1}{c}{(5)}               \\
            &        2SLS         &          0m               &       7-12m         &      13-18m         &      19-24m         \\\hline
            &\multicolumn{5}{c}{Health}                                                                                                         \\\cmidrule(lr){2-6}

Effect of any Medicaid&       0.180\sym{***}&                                         &                     &                     &                     \\
            &    (0.0436)                            &                     &                     &                     &                     \\
[1em]
Complier mean Y&                     &       -0.152\sym{***}        &      0.0785\sym{*}  &     -0.0197         &      -0.114         \\
            &                     &    (0.0366)                &    (0.0360)         &    (0.0242)         &     (0.116)         \\[1em]
            &\multicolumn{5}{c}{Utilization}                                                                                                    \\\cmidrule(lr){2-6}
Effect of any Medicaid&       0.108\sym{**} &                     &                                      &                     &                     \\
            &    (0.0416)         &                                  &                     &                     &                     \\
[1em]
Complier mean Y&                     &      -0.0429          &     -0.0671         &       0.150\sym{***}&       0.171         \\
            &                     &    (0.0327)                &    (0.0415)         &    (0.0261)         &     (0.134)         \\[1em]
\hline
\(N\)       &       17896         &       17896                &       17896         &       17896         &       17896         \\
Share of compliers&                     &          1                &       0.325         &       0.620         &      0.0735         \\
\hline\hline
\end{tabular}
}

%% file: arxiv_mar2024.bib
@article{kowalski2021reconciling,
  title={Reconciling seemingly contradictory results from the Oregon health insurance experiment and the Massachusetts health reform},
  author={Kowalski, Amanda E},
  journal={Review of Economics and Statistics},
  pages={1--45},
  year={2021},
  publisher={MIT Press One Rogers Street, Cambridge, MA 02142-1209, USA journals-info~…}
}

@techreport{blandhol2022tsls,
  title={When is TSLS actually LATE?},
  author={Blandhol, Christine and Bonney, John and Mogstad, Magne and Torgovitsky, Alexander},
  year={2022},
  institution={National Bureau of Economic Research}
}

@article{card2005estimating,
  title={Estimating the effects of a time-limited earnings subsidy for welfare-leavers},
  author={Card, David and Hyslop, Dean R},
  journal={Econometrica},
  volume={73},
  number={6},
  pages={1723--1770},
  year={2005},
  publisher={Wiley Online Library}
}

@article{heckman1974shadow,
  title={Shadow prices, market wages, and labor supply},
  author={Heckman, James},
  journal={Econometrica: journal of the econometric society},
  pages={679--694},
  year={1974},
  publisher={JSTOR}
}

@article{gronau1974wage,
  title={Wage comparisons--A selectivity bias},
  author={Gronau, Reuben},
  journal={Journal of political Economy},
  volume={82},
  number={6},
  pages={1119--1143},
  year={1974},
  publisher={The University of Chicago Press}
}

@article{finkelstein2012,
    author = {Finkelstein, Amy and Taubman, Sarah and Wright, Bill and Bernstein, Mira and Gruber, Jonathan and Newhouse, Joseph P. and Allen, Heidi and Baicker, Katherine and Oregon Health Study Group},
    title = "{The Oregon Health Insurance Experiment: Evidence from the First Year}",
    journal = {The Quarterly Journal of Economics},
    volume = {127},
    number = {3},
    pages = {1057-1106},
    year = {2012},
    issn = {0033-5533},
    doi = {10.1093/qje/qjs020},
    url = {https://doi.org/10.1093/qje/qjs020},
    eprint = {https://academic.oup.com/qje/article-pdf/127/3/1057/30456845/qjs020.pdf},
}

@article{heckman2018unordered,
  title={Unordered monotonicity},
  author={Heckman, James J and Pinto, Rodrigo},
  journal={Econometrica},
  volume={86},
  number={1},
  pages={1--35},
  year={2018}
}

@article{heckman2010JEL,
Author = {Heckman, James J.},
Title = {Building Bridges between Structural and Program Evaluation Approaches to Evaluating Policy},
Journal = {Journal of Economic Literature},
Volume = {48},
Number = {2},
Year = {2010},
Month = {June},
Pages = {356-98},
DOI = {10.1257/jel.48.2.356},
URL = {https://www.aeaweb.org/articles?id=10.1257/jel.48.2.356}}

@article{heckman1999,
 ISSN = {00278424},
 URL = {http://www.jstor.org/stable/47634},
 author = {James J. Heckman and Edward J. Vytlacil},
 journal = {Proceedings of the National Academy of Sciences of the United States of America},
 number = {8},
 pages = {4730-4734},
 publisher = {National Academy of Sciences},
 title = {Local Instrumental Variables and Latent Variable Models for Identifying and Bounding Treatment Effects},
 volume = {96},
 year = {1999}
}

@article{CSS2018,
author = {Chetverikov, Denis and Santos, Andres and Shaikh, Azeem M.},
title = {The Econometrics of Shape Restrictions},
journal = {Annual Review of Economics},
volume = {10},
number = {1},
pages = {31-63},
year = {2018},
doi = {10.1146/annurev-economics-080217-053417},

URL = { 
        https://doi.org/10.1146/annurev-economics-080217-053417
    
},
eprint = { 
        https://doi.org/10.1146/annurev-economics-080217-053417
    
}
}

@article{bai2019practical,
  title={A practical method for testing many moment inequalities},
  author={Bai, Yuehao and Santos, Andres and Shaikh, Azeem},
  journal={University of Chicago, Becker Friedman Institute for Economics Working Paper},
  number={2019-116},
  year={2019}
}

@article{Arteaga2020,
  title={Parental Incarceration and Children's Educational Attainment},
  author={Arteaga, Carolina},
  journal={The Review of Economics and Statistics},
  pages={1--45},
  year={2020}
}

@article{kling2007experimental,
  title={Experimental analysis of neighborhood effects},
  author={Kling, Jeffrey R and Liebman, Jeffrey B and Katz, Lawrence F},
  journal={Econometrica},
  volume={75},
  number={1},
  pages={83--119},
  year={2007},
  publisher={Wiley Online Library}
}

@article{heller2017thinking,
  title={Thinking, fast and slow? Some field experiments to reduce crime and dropout in Chicago},
  author={Heller, Sara B and Shah, Anuj K and Guryan, Jonathan and Ludwig, Jens and Mullainathan, Sendhil and Pollack, Harold A},
  journal={The Quarterly Journal of Economics},
  volume={132},
  number={1},
  pages={1--54},
  month = {10},
  year={2017}
}

@article{katz2001moving,
  title={Moving to opportunity in Boston: Early results of a randomized mobility experiment},
  author={Katz, Lawrence F and Kling, Jeffrey R and Liebman, Jeffrey B},
  journal={The Quarterly Journal of Economics},
  volume={116},
  number={2},
  pages={607--654},
  year={2001},
  publisher={MIT Press}
}

@article{Norris2020,
  title={The effects of parental and sibling incarceration: Evidence from ohio},
  author={Norris, Samuel and Pecenco, Matthew and Weaver, Jeffrey},
  journal={American Economic Review},
  volume={111},
  number={9},
  pages={2926--63},
  year={2021}
}

@article{abadie2003semiparametric,
  title={Semiparametric instrumental variable estimation of treatment response models},
  author={Abadie, Alberto},
  journal={Journal of econometrics},
  volume={113},
  number={2},
  pages={231--263},
  year={2003},
  publisher={Elsevier}
}

@article{hong2020numerical,
  title={The numerical bootstrap},
  author={Hong, Han and Li, Jessie and others},
  journal={The Annals of Statistics},
  volume={48},
  number={1},
  pages={397--412},
  year={2020},
  publisher={Institute of Mathematical Statistics}
}

@article{fang2019inference,
  title={Inference on directionally differentiable functions},
  author={Fang, Zheng and Santos, Andres},
  journal={The Review of Economic Studies},
  volume={86},
  number={1},
  pages={377--412},
  year={2019},
  publisher={Oxford University Press}
}

@techreport{andrews2019inference,
  title={Inference for linear conditional moment inequalities},
  author={Andrews, Isaiah and Roth, Jonathan and Pakes, Ariel},
  year={2019},
  institution={National Bureau of Economic Research}
}

@article{aizer2015,
author = {Aizer, Anna and Doyle, Joseph J.}, 
title = {Juvenile Incarceration, Human Capital, and Future Crime: Evidence from Randomly Assigned Judges},
volume={130},
number={2},
year = {2015}, 
pages={759-803},
journal = {The Quarterly Journal of Economics} 
}

@article{andrews2012inference,
  title={Inference for parameters defined by moment inequalities: A recommended moment selection procedure},
  author={Andrews, Donald W. and Barwick, Panle Jia},
  journal={Econometrica},
  volume={80},
  number={6},
  pages={2805--2826},
  year={2012},
  publisher={Wiley Online Library}
}

@article{andrews2013inference,
  title={Inference based on conditional moment inequalities},
  author={Andrews, Donald W. and Shi, Xiaoxia},
  journal={Econometrica},
  volume={81},
  number={2},
  pages={609--666},
  year={2013},
  publisher={Wiley Online Library}
}

@article{cox2020simple,
  title={Simple Adaptive Size-Exact Testing for Full-Vector and Subvector Inference in Moment Inequality Models},
  author={Cox, Gregory and Shi, Xiaoxia},
  journal={arXiv preprint arXiv:1907.06317},
  year={2020}
}

@article{andrews2010inference,
  title={Inference for parameters defined by moment inequalities using generalized moment selection},
  author={Andrews, Donald WK and Soares, Gustavo},
  journal={Econometrica},
  volume={78},
  number={1},
  pages={119--157},
  year={2010},
  publisher={Wiley Online Library}
}

@inproceedings{canay2017practical,
  title={Practical and theoretical advances in inference for partially identified models},
  author={Canay, Ivan A. and Shaikh, Azeem M.},
  booktitle={Advances in Economics and Econometrics: Eleventh World Congress},
  volume={2},
  pages={271--306},
  year={2017},
  organization={Cambridge University Press Cambridge}
}

@article{norris2019examiner,
  title={Examiner inconsistency: Evidence from refugee appeals},
  author={Norris, Samuel},
  journal={University of Chicago, Becker Friedman Institute for Economics Working Paper},
  number={2018-75},
  year={2019}
}

@techreport{frandsen2019judging,
  title={Judging judge fixed effects},
  author={Frandsen, Brigham R and Lefgren, Lars J. and Leslie, Emily C.},
  year={2019},
  institution={National Bureau of Economic Research}
}

@article{chernozhukov2015constrained,
  title={Constrained conditional moment restriction models},
  author={Chernozhukov, Victor and Newey, Whitney K and Santos, Andres},
  year={2020}
}

@article{mogstad2018using,
  title={Using instrumental variables for inference about policy relevant treatment parameters},
  author={Mogstad, Magne and Santos, Andres and Torgovitsky, Alexander},
  journal={Econometrica},
  volume={86},
  number={5},
  pages={1589--1619},
  year={2018},
  publisher={Wiley Online Library}
}

@article{RSW2014,
  title={A practical two-step method for testing moment inequalities},
  author={Romano, Joseph P and Shaikh, Azeem M and Wolf, Michael},
  journal={Econometrica},
  volume={82},
  number={5},
  pages={1979--2002},
  year={2014}
}

@article{CCK2018,
    author = {Chernozhukov, Victor and Chetverikov, Denis and Kato, Kengo},
    title = "{Inference on Causal and Structural Parameters using Many Moment Inequalities}",
    journal = {The Review of Economic Studies},
    volume = {86},
    number = {5},
    pages = {1867-1900},
    year = {2018},
    month = {11},
    issn = {0034-6527},
    doi = {10.1093/restud/rdy065},
    url = {https://doi.org/10.1093/restud/rdy065},
    eprint = {https://academic.oup.com/restud/article-pdf/86/5/1867/29581443/rdy065.pdf}
}

@article{marx2020sharp,
  title={Sharp Bounds in the Latent Index Selection Model},
  author={Marx, Philip},
  journal={arXiv preprint arXiv:2012.02390},
  year={2020}
}

@article{marshall2016coarsening,
  title={Coarsening bias: How coarse treatment measurement upwardly biases instrumental variable estimates},
  author={Marshall, John},
  journal={Political Analysis},
  volume={24},
  number={2},
  pages={157--171},
  year={2016},
  publisher={Cambridge University Press}
}

@article{Huber_Mellace2015,
author = {Huber, Martin and Mellace, Giovanni},
title = {Testing Instrument Validity for LATE Identification Based on Inequality Moment Constraints},
journal = {The Review of Economics and Statistics},
volume = {97},
number = {2},
pages = {398-411},
year = {2015}
}

@article{angrist1991does,
  title={Does compulsory school attendance affect schooling and earnings?},
  author={Angrist, Joshua D. and Krueger, Alan B.},
  journal={The Quarterly Journal of Economics},
  volume={106},
  number={4},
  pages={979--1014},
  year={1991},
  publisher={MIT Press}
}

@article{heckman_etal2006,
 ISSN = {00346535, 15309142},
 URL = {http://www.jstor.org/stable/40043006},
 author = {James J. Heckman and Sergio Urzua and Edward J. Vytlacil},
 journal = {The Review of Economics and Statistics},
 number = {3},
 pages = {389-432},
 publisher = {The MIT Press},
 title = {Understanding Instrumental Variables in Models with Essential Heterogeneity},
 volume = {88},
 year = {2006}
}

@article{heckman2005,
 ISSN = {00129682, 14680262},
 URL = {http://www.jstor.org/stable/3598865},
 author = {James J. Heckman and Edward J. Vytlacil},
 journal = {Econometrica},
 number = {3},
 pages = {669-738},
 publisher = {[Wiley, Econometric Society]},
 title = {Structural Equations, Treatment Effects, and Econometric Policy Evaluation},
 volume = {73},
 year = {2005}
}

@book{deaton1980economics,
  title={Economics and consumer behavior},
  author={Deaton, Angus and Muellbauer, John},
  year={1980},
  publisher={Cambridge university press}
}

@article{balkepearl97,
	author = { Alexander   Balke  and  Judea   Pearl },
	title = {Bounds on Treatment Effects from Studies with Imperfect Compliance},
	journal = {Journal of the American Statistical Association},
	volume = {92},
	number = {439},
	pages = {1171-1176},
	year  = {1997},
	publisher = {Taylor & Francis},
	doi = {10.1080/01621459.1997.10474074},
	
	URL = { 
	https://doi.org/10.1080/01621459.1997.10474074
	
	},
	eprint = { 
	https://doi.org/10.1080/01621459.1997.10474074
	
	}
}

@article{vytlacil2002,
author = {Edward Vytlacil},
title = {Independence, Monotonicity, and Latent Index Models: An Equivalence Result},
journal = {Econometrica},
volume = {70},
number = {1},
year = {2002},
pages = {331-341},
doi = {10.1111/1468-0262.00277},
url = {https://onlinelibrary.wiley.com/doi/abs/10.1111/1468-0262.00277},
eprint = {https://onlinelibrary.wiley.com/doi/pdf/10.1111/1468-0262.00277}
}

@article{vytlacil2006,
author = { Edward  Vytlacil },
title = {Ordered Discrete-Choice Selection Models and Local Average Treatment Effect Assumptions: Equivalence, Nonequivalence, and Representation Results},
journal = {The Review of Economics and Statistics},
volume = {88},
number = {3},
pages = {578-581},
year = {2006},
doi = {10.1162/rest.88.3.578},

URL = { 
        https://doi.org/10.1162/rest.88.3.578
    
},
eprint = { 
        https://doi.org/10.1162/rest.88.3.578
    
}
}

@article{mourifie2017testing,
  title={Testing local average treatment effect assumptions},
  author={Mourifi{\'e}, Ismael and Wan, Yuanyuan},
  journal={Review of Economics and Statistics},
  volume={99},
  number={2},
  pages={305--313},
  year={2017}
}

@article{eckhoff2018instrument,
    author = {Andresen, Martin E and Huber, Martin},
    title = {Instrument-based estimation with binarised treatments: issues and tests for the exclusion restriction},
    journal = {The Econometrics Journal},
    year = {2021}
}

@article{angrist1995,
author = {Joshua D.   Angrist  and  Guido W.   Imbens },
title = {Two-Stage Least Squares Estimation of Average Causal Effects in Models with Variable Treatment Intensity},
journal = {Journal of the American Statistical Association},
volume = {90},
number = {430},
pages = {431-442},
year  = {1995},
publisher = {Taylor & Francis},
doi = {10.1080/01621459.1995.10476535},
URL = { 
        https://amstat.tandfonline.com/doi/abs/10.1080/01621459.1995.10476535
    
},
eprint = { 
        https://amstat.tandfonline.com/doi/pdf/10.1080/01621459.1995.10476535
    
}
}

@article{mulligan2008selection,
  title={Selection, investment, and women's relative wages over time},
  author={Mulligan, Casey B and Rubinstein, Yona},
  journal={The Quarterly Journal of Economics},
  volume={123},
  number={3},
  pages={1061--1110},
  year={2008},
  publisher={MIT Press}
}

@article{Dahl2002,
author = {Dahl, Gordon B.},
title = {Mobility and the Return to Education: Testing a Roy Model with Multiple Markets},
journal = {Econometrica},
volume = {70},
number = {6},
pages = {2367-2420},
year = {2002}
}

@article{rose2018does,
  title={Does incarceration increase crime?},
  author={Rose, Evan K and Shem-Tov, Yotam},
  journal={Available at SSRN 3205613},
  year={2018}
}

@article{rose_shemtov2019does,
  title={How does incarceration affect reoffending? Estimating the dose-response function},
  author={Rose, Evan K and Shem-Tov, Yotam},
  journal={Journal of Political Economy},
  volume={129},
  number={12},
  pages={3302--3356},
  year={2021},
  publisher={The University of Chicago Press Chicago, IL}
}

@article{rose_shemtov2019WIP,
  title={Does Incarceration Increase Crime?},
  author={Rose, Evan K. and Shem-Tov, Yotam},
  journal={Working Paper},
  year={2019}
}

@article{goldin2019health,
    author = {Goldin, Jacob and Lurie, Ithai Z and McCubbin, Janet},
    title = "{Health Insurance and Mortality: Experimental Evidence from Taxpayer Outreach}",
    journal = {The Quarterly Journal of Economics},
    volume = {136},
    number = {1},
    pages = {1-49},
    year = {2020},
    month = {09},
    issn = {0033-5533},
    doi = {10.1093/qje/qjaa029},
    url = {https://doi.org/10.1093/qje/qjaa029},
    eprint = {https://academic.oup.com/qje/article-pdf/136/1/1/35073172/qjaa029.pdf},
}

@article{Bhuller_etal2018,
	author = {Bhuller, Manudeep and Dahl, Gordon B. and L{\o}ken, Katrine V. and Mogstad, Magne},
	title = {Incarceration, Recidivism, and Employment},
	journal = {Journal of Political Economy},
	volume = {128},
	number = {4},
	pages = {1269-1324},
	year = {2020},
	doi = {10.1086/705330},
	
	URL = { 
	https://doi.org/10.1086/705330
	
	},
	eprint = { 
	https://doi.org/10.1086/705330
	
	}
}

@article{imbens1997,
author = {Imbens, Guido and Rubin, Donald}, 
title = {Estimating Outcome Distributions for Compliers in Instrumental Variables Models},
volume = {64}, 
number = {4}, 
pages = {555-574}, 
year = {1997}, 
journal = {The Review of Economic Studies} 
}

@article{imbens1994,
 author = {Guido W. Imbens and Joshua D. Angrist},
 journal = {Econometrica},
 number = {2},
 pages = {467-475},
 publisher = {The Econometric Society},
 title = {Identification and Estimation of Local Average Treatment Effects},
 volume = {62},
 year = {1994}
}

@inproceedings{card1995,
  title={Using Geographic Variation in College Proximity to Estimate the Return to Schooling},
  author={Card, David},
  booktitle={Aspects of Labor Market Behaviour: Essays in Honour of John Vanderkamp},
  pages={201-222},
  year={1995},
  organization={University of Toronto Press, Toronto}
}

@article{kitagawa2015test,
  title={A test for instrument validity},
  author={Kitagawa, Toru},
  journal={Econometrica},
  volume={83},
  number={5},
  pages={2043--2063},
  year={2015},
  publisher={Wiley Online Library}
}
